\documentstyle[11pt,aaspp4]{article}

\lefthead{MCDAVID}

\righthead{MULTICOLOR POLARIMETRY OF Be STARS}

\begin{document}

\title{Multicolor Polarimetry of Selected Be Stars: 1995--98}

\author{David McDavid}

\affil{Limber Observatory, Timber Creek Road, P.O. Box 63599, Pipe Creek TX 78063-3599; \\ mcdavid@limber.org}

\authoraddr{Timber Creek Road, P.O. Box 63599, Pipe Creek TX 78063-3599}

\begin{abstract}

A new polarimeter called AnyPol has been used at Limber Observatory
for four years to annually monitor the broadband linear polarization
of a sample of bright northern Be stars.  This is the fourth report on
a program started in 1985 at McDonald Observatory and the first one to
come entirely from the new installation.  Although no variability was
detected at the 3$\sigma$ level during the current reporting period,
analysis of the full 13-year data set is beginning to reveal hints of
long-term variability that may provide clues for understanding the Be
phenomenon.


\end{abstract}

\section{INTRODUCTION}

Be stars are non-supergiant B-type stars whose spectra have, or had at
some time, one or more Balmer lines in emission.  The mystery of the
Be phenomenon is that the emission, which is well understood to
originate from a flattened circumstellar envelope or disk, can come
and go episodically on time scales of days to decades.  This has yet
to be explained as a predictable consequence of stellar evolution
theory, although many contributing factors have been discussed,
including rapid rotation, radiation-driven stellar winds, nonradial
pulsation, flarelike magnetic activity, and binary interaction.  For
the unfamiliar reader, the review of Be stars by Slettebak
\markcite{Slet1} (1988) will provide an excellent introduction.

Recent optical interferometry combined with spectropolarimetry has
directly confirmed that the circumstellar envelopes surrounding Be
stars are equatorially flattened \markcite{Q1} (Quirrenbach et
al. 1997).  Given the small observed polarization of only about 1\%,
Monte Carlo computer simulations of polarization by electron
scattering of the starlight in the circumstellar envelope
\markcite{WBB1} (Wood, Bjorkman, \& Bjorkman 1997) constrain it to be
an extremely thin disk, with opening angle (half width in latitude) on
the order of 3$\arcdeg$.  This example shows that polarimetry is a
very useful observational technique for studying physical properties
of the envelopes, with the ultimate purpose of understanding their
origin.  Toward this end it is of great interest to measure
polarization variations and the time scales on which they occur, in
order to characterize the physical processes involved.

This paper is the fourth status report on an ongoing program of annual
polarimetric monitoring of a sample of bright northern Be stars begun
in 1985 \markcite{DM1} (McDavid 1994 and references therein).  The
program stars are listed in Table 1, with visual magnitudes taken from
Hoffleit and Jaschek \markcite{HJ1} (1982) and spectral types and
$v \sin i$ values from Slettebak \markcite{Slet2} (1982).  A broadband
filter system (Table 2) was chosen to extend the time base of
continuous systematic observations for the study of long-term
variability, taking advantage of earlier work which dates back to the
1950s.  After 10 years on the 0.9 m telescope at McDonald Observatory
the project was relocated to a new polarimeter on the 0.4 m telescope
at Limber Observatory, where more flexible scheduling allows the study
of variability on a greater variety of time scales.  Advancements in
commercially available instrumentation have made it possible to do so
without compromising the quality of the data.

\section{ANYPOL: A GENERIC LINEAR POLARIMETER}

AnyPol got its name from the fact that it is completely generic and
incorporates no new principles of design or construction. In most
respects, including the color system as specified by prescription, it
is a simplified and miniaturized version of the McDonald Observatory
polarimeter \markcite{Bre1} (Breger 1979): a rapidly rotating
Glan-Taylor prism as an analyzer, followed by a Lyot depolarizer,
Johnson/Cousins $\ubvr I$ glass filters, a Fabry lens, and an uncooled
S-20 photomultiplier tube.  It is very compact, with stepper motors
and belt drives for separate analyzer and filter wheel modules which
fit into a single main head unit. A simple postviewer using a
right-angle prism at one position of the filter wheel is adequate for
finding and centering, and three aperture sizes are available in a
slide mechanism with a pair of LEDs for backlighting.

The control system for AnyPol is based on a 66 MHz 486 PC with plugin
multichannel analyzer and stepper motor controller cards, interfaced
to the polarimeter head through an electronics chassis unit containing
the power supplies and microstepping drivers. A point-and-click
display serves to control all selectable functions and parameters and
also shows 10 s updates of the measurement in progress, including a
graph of the data buffer. The mathematical details of the data
processing were adapted from the control program for the Minipol
polarimeter of the University of Arizona \markcite{FS1} (Frecker \&
Serkowski 1976).

On the 0.4 m telescope at Limber Observatory the photon count rates
are comparable to those obtained with the 0.9 m telescope at McDonald
Observatory with a neutral density filter of 10\% transmission which
was necessary for the bright stars ($V=2$--5) in the Be star
monitoring program.  Observational error estimates are derived from
the repeatability of multiple independent measurements, and they are
in general agreement with checks based on the residuals in the fit to
the modulated signal and the uncertainty derived from photon counting
statistics.  Typical errors are on the order of 0.05\% in the degree
of polarization and 2$\arcdeg$ in position angle.  One of the main
sources of error is a slight variation in the speed of the motor
driving the analyzer, which becomes significant at the level of a few
hundredths of a percent.  All observations are corrected for an
instrumental polarization on the order of 0.10\%, tracked by repeated
observations of unpolarized standard stars from the list of
\markcite{Serk1} Serkowski (1974).  The position angle is calibrated
by observing polarized standard stars from the list of \markcite{HB1}
Hsu \& Breger (1982).

\section{OBSERVATIONS}

The targets for the annual monitoring program were selected to include
Be stars with a variety of different characteristics and with the
longest possible history of continuous observation.  They fall into
summer and winter groups, so the basic observing strategy is to make
the polarization measurements during about one week in summer and one
week in winter.  The only interruption in the project has been the
loss of 1994 while the new polarimeter was under construction.

One individual measurement consists of 3 cycles through all 5 filters
with a 200 s integration time on each filter.  If there is a bright
Moon, a sky cycle is taken to correct for the background polarization.
The result for each single filter is taken to be the mean and standard
deviation of the 3 integrations in that filter.  (The standard
deviation is a more conservative error estimate than the standard
deviation of the mean, but it may be more realistic because 3
measurements is a very small sample.)  During a typical observing run
for this project, about 3 to 5 observations of each program star are
collected on different nights during the same week.

\section{ANALYSIS OF VARIABILITY}

The first goal in analyzing the data is to identify clear cases of
variable polarization over the time scales covered by this installment
of the project.  The data are presented in Tables 3--12, which begin
with the month and year of the observing run and the number of
measurements in each filter.  The $q$ and $u$ normalized Stokes
parameters, the degree of polarization $p$, and the polarization
position angle $\theta$ are all given as the mean and standard
deviation for each run.  The last column shows the average error in
$p$ and $\theta$ for a single measurement.  In addition to the program
Be stars, two polarized standard stars were also observed as checks on
the stability of the system: 2H Cam (HD 21291, HR 1035, $V=4.21$, B9
Ia) in winter and o Sco (HD 147084, HR 6081, $V=4.54$, A5 II) in
summer.

As in previous papers in this series, summary tables (Tables 13--15)
were constructed giving the means and standard deviations of the
measurements for each star in each filter over the four annual data
sets.  The quantities in angled brackets are also four-year averages,
and the quantities in rows labeled ``GAV'' for ``grand average'' are
averages over all five filters.

Since $p$ and $\theta$ carry a statistical bias, $q$ and $u$ are more
appropriate quantities for evaluating variability \markcite{CS}
(Clarke \& Stewart 1986).  Nevertheless, it may provide more physical
insight to study $p$ and $\theta$ to see if the variability is mainly
in polarization degree or position angle.  In fact, with the limited
number of measurements at hand, statistical tests must be applied with
caution regardless of which set of parameters is used.

We can apply various simple 3$\sigma$ criteria to conservatively
identify variability, as in previous papers in this series.  Looking
for night-to-night variability, we see that Tables 3--12 show only two
cases in which $dq$ or $du$ is greater than 3$dpi$: $du^{B}$ of
$\gamma$ Cas in 01/96 and $du^{B}$ of o And in 06/95.  Both cases are
negligible, since they occur in only one filter and during only one
observing run.  For year-to-year variability, Tables 13--15 show not a
single case in which $dq$ or $du$ is greater than 3{\tt <}$dpi${\tt
>}.  The conclusion is that we can demonstrate no statistically
significant variability in the polarization of any of the program
stars over the latest 4-year time period.

Since variable polarization was detected in the two previous reports
on this project, it seems advisable to search for an explanation by
comparing the Limber Observatory system with the system used at
McDonald Observatory.  For this purpose Table 13 of \markcite{DM1}
McDavid (1994) is reproduced here as Table 16 for direct comparison
with the present Table 15.  These observations of polarized standard
stars show very clearly that the two systems match extremely well.
The only outstanding difference is the typical precision of a single
observation, which is higher in the McDonald system.  This is readily
understood since the McDonald estimates were based on theoretical
photon counting statistics, while the Limber estimates are based on
experimental scatter in repeated measurements.  With a larger value
for the error in a single observation, the Limber system is sometimes
a less sensitive detector of variability, but it may also give more
realistic results.

Work is underway to make a complete data set available at the
Strasbourg astronomical Data Center (CDS), including all of the annual
observations from both McDonald Observatory and Limber Observatory
published to date in this series of papers.  The times will be given
in decimal years for better precision than the current specification
of month and year.

\section{DISCUSSION}

The 3$\sigma$ criterion used here and in the previous papers of this
series is very conservative, practically guaranteeing the validity of
any detections of variability.  This project has shown that such
unquestionable detections are by no means common.  However, as the
data base has now grown to cover more than a decade in time, some
patterns of long-term variability are beginning to appear, even though
they may not have been previously recognizable at the 3$\sigma$ level.
What follows is a commentary on the behavior of each individual
program star, illustrated with q-u plots and graphs of intrinsic
polarization as a function of time in all 5 filters over the entire
duration of the monitoring program.

In each q-u plot the data points are filled circles, the mean is a
cross drawn to the size of the average error of a single measurement,
the standard deviation is represented by a dotted ellipse centered on
the mean, and three times the average error of a single measurement is
represented by a solid ellipse centered on the mean.  For each star
there is one additional q-u plot showing the mean value for each
filter.  Note that for all the program stars except 48 Lib it is
possible to fit the 5 single-filter data points with a straight line
which passes close to the origin.  This implies that either the
interstellar component of the polarization is small or its position
angle is nearly the same as that of the intrinsic component.  In
either case, the straight line fit gives a good approximation to the
position angle of the intrinsic polarization.  Any elongation of the
distribution of data points along that general direction in the
single-filter q-u plots is good evidence for intrinsic polarization
that is variable in degree but constant in position angle, as is
commonly expected for polarization caused by electron scaterring in an
equatorially flattened axisymmetric disk.  This technique makes it
possible to identify variable intrinsic polarization even when it is
too small to meet the 3$\sigma$ test.

The graphs of polarization degree and position angle as a function of
time for each filter show the intrinsic polarization, calculated by
vectorially subtracting the interstellar component determined by
\markcite{PBL1} Poeckert, Bastien, \& Landstreet (1979) or by
\markcite{MB1} McLean \& Brown (1978), using a Serkowski law of the
form $p_{ISi} = p_{max} \exp [-1.15 \ln^{2} (\lambda_{max} /
\lambda_{i})]$ with parameters as summarized in Table 17.

\subsection{Gamma Cas}

Gamma Cas provides a good example of how the mean values of the
polarization in 5 filters can sometimes be nearly collinear in the q-u
plane, so that a straight line fit can give a good approximation to
the position angle of the intrinsic polarization (see Figure 1, lower
right panel).  The individual filter plots all show some evidence for
elongation of the data point patterns along this direction (especially
in $U$ and $R$), which is good evidence that there is some real
low-level variability.  Figure 2, however, shows only slight changes
from year to year.  The polarization of $\gamma$ Cas, and therefore
the state of its circumstellar envelope, appears to have been mostly
stable since this monitoring program began.

\subsection{Phi Per}

The polarization of $\phi$ Per has been almost certainly variable from
year to year, as may be seen in Figure 3, where the data patterns are
clearly elongated along the direction indicated by a straight line fit
to the filter means.  Some sinusoidal tendencies can be seen in Figure
4, especially in the $R$ bandpass.  Periodogram analysis suggests a
period on the order of 11 to 12 years.  If the circumstellar disk is
tilted with respect to the binary orbital plane, as advanced by
\markcite{CB1} Clarke \& Bjorkman (1998), it might be expected to
precess with a similar period.  However, the almost perfectly constant
position angle of the polarization argues against this explanation for
the periodic polarization.

\subsection{48 Per}

With $v \sin i$ = 200 km ${\tt s^{-1}}$, 48 Per is probably viewed at
a relatively low inclination of its rotation axis to the line of
sight.  This is consistent with its relatively small polarization,
even though it is a strong H$\alpha$ emitter.  It is interesting to
see from Figure 6 that the position angle shows stronger variability
than the degree of polarization, including a hint of a 4--5 year cycle
in the $B$ filter.  A precessing bar-shaped nonuniformity embedded in
the disk would be expected to generate this kind of variability.  The
absence of a preferred direction in the q-u plots (Figure 5) lends
further strength to this interpretation.

\subsection{Zeta Tau}

Zeta Tau is one of the most highly polarized and strongly variable of
all the program stars.  A look at Figure 8 shows a slow and steady
rise in the degree of polarization with occasional mild outbursts or
local maxima, while the position angle remains constant.  The filter
averages in Figure 7 are very nearly collinear, and the individual
filter plots clearly show alignment in that direction.

Work is in progress on a possible correlation between continuum
polarization and $V/R$ variations of the H$\alpha$ emission line
profile of $\zeta$ Tau as a test of the theory of ``one-armed''
density perturbations of the circumstellar disk \markcite{AO1}
(Okazaki 1997) and their effects on the polarization.  Hopefully the
results will place some constraints on the nature of Be disks and the
processes leading to their formation.

\subsection{48 Lib}

In Figure 9 the intrinsic position angle of the polarization of 48 Lib
is poorly determined because the interstellar component is large and
has a very different position angle than the intrinsic component.  In
Figure 10 the interstellar polarization has been removed, resulting in
a better straight line fit that passes acceptably close to the origin
to approximate the intrinsic position angle.  This angle is indeed
favored by the elongations of the q-u data sets in the individual
filters.

Figure 11 shows that the degree of intrinsic polarization is nearly
cyclic with a period on the order of 4--5 years, although somewhat
noisy due to short-term variations.  As in the case of $\zeta$ Tau
there is a correlation between the polarization period and that of
$V/R$ in the H$\alpha$ emission line, and it is currently being
pursued in the same context.

\subsection{Chi Oph}

Chi Oph has the lowest $v \sin i$ of all the program stars: 140 km
${\tt s^{-1}}$.  It also has one of the least degrees of intrinsic
polarization, as would be expected if its rotation axis is only
slightly inclined to the line of sight.  In Figure 12 the fit of the
intrinsic position angle line is somewhat weak, and the individual
filter plots have only a slight tendency for alignment.

The graphs of Figure 13 show how small the intrinsic polarization is
and how poorly determined the position angle is as a result.  Apart
from an interesting low-amplitude ripple in the degree of
polarization, there is little evidence for any significant
variability.

\subsection{Pi Aqr}

Figure 14 shows what $>3\sigma$ variable polarization of a Be star
should look like in the q-u plane.  The degree of intrinsic
polarization of $\pi$ Aqr has changed more by far than that of any
other star on the program list.  The strong alignment evident in the
q-u plots indicates that the position angle is very stable.  This is
also demonstrated by the position angle graphs of Figure 15.

When monitoring began in 1985, $\pi$ Aqr had nearly the largest (if
not the largest) polarization of any Be star in the sky.  Now, 13
years later, that polarization has almost completely disappeared.  If
we could explain exactly what happened, we would be close to
understanding the Be phenomenon itself.  Did a dynamic disk lose a
source of continuous replenishment?  If there were such a source, what
might have turned it off?  Was the disk a static structure?  If so,
what prompted it to dissipate?  Were there any changes in the
underlying star?  There are still far more questions than answers.

\subsection{Omicron And}

Since 1986 the polarization of o And has been increasing almost
uniformly except for two or three minor outbursts and dropouts.  The
relatively small degree of polarization makes the position angle
difficult to determine (see Figure 16), but Figure 17 shows good
definition of the intrinsic position angle based on fitting to the
filter means.  The polarization is well-behaved in the single filter
q-u plots, which show that there is real variability by their strong
alignment to the intrinsic position angle.  This gives a good record
of the gradual buildup of a polarizing Be envelope or disk.

\acknowledgements

I am very grateful to Michel Breger and Santiago Tapia for introducing
me to the basics of astronomical polarimetry. I also thank Paul
Krueger for his skillful machine work in transforming AnyPol from line
drawings on paper into the reality of metal.  Jon Bjorkman's
constructive suggestions as referee helped to clarify this
presentation in many ways.

\newpage

Fig.1.-Normalized Stokes parameter plots of the polarization
of $\gamma$ Cas (see text for explanation of the symbols).

Fig.2.-Degree and position angle of the intrinsic polarization
of $\gamma$ Cas.

Fig.3.-Normalized Stokes parameter plots of the polarization
of $\phi$ Per (see text for explanation of the symbols).

Fig.4.-Degree and position angle of the intrinsic polarization
of $\phi$ Per.

Fig.5.-Normalized Stokes parameter plots of the polarization
of 48 Per (see text for explanation of the symbols).

Fig.6.-Degree and position angle of the intrinsic polarization
of 48 Per.

Fig.7.-Normalized Stokes parameter plots of the polarization
of $\zeta$ Tau (see text for explanation of the symbols).

Fig.8.-Degree and position angle of the intrinsic polarization
of $\zeta$ Tau.

Fig.9.-Normalized Stokes parameter plots of the polarization
of 48 Lib (see text for explanation of the symbols).

Fig.10.-Normalized Stokes parameter plots of the intrinsic
polarization of 48 Lib (see Subsection 5.5 for comment).

Fig.11.-Degree and position angle of the intrinsic polarization
of 48 Lib.

Fig.12.-Normalized Stokes parameter plots of the polarization
of $\chi$ Oph (see text for explanation of the symbols).

Fig.13.-Degree and position angle of the intrinsic polarization
of $\chi$ Oph.

Fig.14.-Normalized Stokes parameter plots of the polarization
of $\pi$ Aqr (see text for explanation of the symbols).

Fig.15.-Degree and position angle of the intrinsic polarization
of $\pi$ Aqr.

Fig.16.-Normalized Stokes parameter plots of the polarization
of o And (see text for explanation of the symbols).

Fig.17.-Degree and position angle of the intrinsic polarization
of o And.

\newpage

\begin{deluxetable}{lrrrlc}
\tt
\tablecolumns{6}
\tablewidth{0pt}
\tablecaption{Program Be Stars}
\tablehead{
\colhead{Name} &
\colhead{HD} &
\colhead{HR} &
\colhead{V} &
\colhead{Spectral Type} &
\colhead{$v \sin i$} \nl
& & & & & (km s$^{-1}$)
}

\startdata

$\gamma$ Cas & 5394 & 264 & 2.47 & B0.5 IVe & 230 \nl

$\phi$ Per & 10516 & 496 & 4.07 & B1.5 (V:)e-shell & 400 \nl

48 Per & 25940 & 1273 & 4.04 & B4 Ve & 200 \nl

$\zeta$ Tau & 37202 & 1910 & 3.00 & B1 IVe-shell & 220 \nl

48 Lib & 142983 & 5941 & 4.88 & B3:IV:e-shell & 400 \nl

$\chi$ Oph & 148184 & 6118 & 4.42 & B1.5 Ve & 140 \nl

$\pi$ Aqr & 212571 & 8539 & 4.66 & B1 III-IVe & 300 \nl

o And & 217675 & 8762 & 3.62 & B6 III & 260 \nl

\enddata
\end{deluxetable}

\clearpage

\begin{deluxetable}{ccr}
\tt
\tablecolumns{3}
\tablewidth{0pt}
\tablecaption{Filter System Parameters}
\tablehead{
\colhead{ \ \ \ Filter} &
\colhead{Effective Wavelength} &
\colhead{Bandpass (fwhm)} \nl
& (\AA) & (\AA) \ \ \ \ \
}

\startdata

\ \ \ U & 3650 & 700 \ \ \ \ \ \ \nl
\ \ \ B & 4400 & 1000 \ \ \ \ \ \ \nl
\ \ \ V & 5500 & 900 \ \ \ \ \ \ \nl
\ \ \ R & 6400 & 1500 \ \ \ \ \ \ \nl
\ \ \ I & 7900 & 1500 \ \ \ \ \ \ \nl

\enddata
\end{deluxetable}

\clearpage

\begin{deluxetable}{lllccc}
\footnotesize
\tt
\tablecolumns{6}
\tablewidth{0pt}
\tablecaption{2H Cam}
\tablehead{
\colhead{Date(Mo/Yr)} &
\colhead{$q\/$/$dq$} &
\colhead{$u\/$/$du$} &
\colhead{$p\/$/$dp$} &
\colhead{$\theta\/$/$d\theta$} &
\colhead{$dpi\/$/$d\theta i$} \nl
\colhead{Filter($n$)} &
\colhead{(\%)} &
\colhead{(\%)} &
\colhead{(\%)} &
\colhead{($\arcdeg$)} &
\colhead{(\%) ($\arcdeg$)}
}

\startdata
\sidehead{12/94}

U(4) &  -1.82/0.07 \ &  -2.33/0.06 \ & 2.96/0.02 &  116.0/\ 0.8 & 0.09/\ 1.3 \nl

B(4) &  -2.06/0.03 \ &  -2.58/0.09 \ & 3.30/0.08 &  115.7/\ 0.5 & 0.05/\ 0.8 \nl

V(4) &  -2.07/0.08 \ &  -2.69/0.11 \ & 3.40/0.10 &  116.2/\ 0.8 & 0.05/\ 0.8 \nl

R(4) &  -2.00/0.07 \ &  -2.61/0.11 \ & 3.29/0.07 &  116.3/\ 1.0 & 0.04/\ 0.7 \nl

I(4) &  -1.82/0.09 \ &  -2.24/0.02 \ & 2.88/0.06 &  115.5/\ 0.7 & 0.08/\ 0.7 \nl \sidehead{01/96}

U(3) &  -1.94/0.01 \ &  -2.43/0.08 \ & 3.11/0.05 &  115.7/\ 0.6 & 0.10/\ 1.0 \nl

B(3) &  -1.97/0.11 \ &  -2.67/0.04 \ & 3.32/0.10 &  116.8/\ 0.6 & 0.08/\ 0.7 \nl

V(3) &  -2.06/0.09 \ &  -2.75/0.05 \ & 3.44/0.09 &  116.6/\ 0.4 & 0.04/\ 0.5 \nl

R(3) &  -1.98/0.06 \ &  -2.70/0.08 \ & 3.35/0.10 &  116.9/\ 0.3 & 0.06/\ 0.3 \nl

I(3) &  -1.79/0.10 \ &  -2.43/0.04 \ & 3.02/0.09 &  116.8/\ 0.6 & 0.11/\ 0.9 \nl
\sidehead{12/96}

U(3) &  -1.73/0.02 \ &  -2.44/0.02 \ & 3.00/0.02 &  117.4/\ 0.2 & 0.06/\ 0.9 \nl

B(3) &  -1.86/0.10 \ &  -2.72/0.03 \ & 3.29/0.05 &  117.8/\ 0.8 & 0.07/\ 0.3 \nl

V(3) &  -2.03/0.06 \ &  -2.78/0.05 \ & 3.45/0.01 &  117.0/\ 0.6 & 0.04/\ 0.5 \nl

R(3) &  -1.98/0.05 \ &  -2.66/0.03 \ & 3.32/0.01 &  116.7/\ 0.5 & 0.05/\ 0.4 \nl

I(3) &  -1.80/0.01 \ &  -2.41/0.08 \ & 3.01/0.07 &  116.6/\ 0.5 & 0.10/\ 0.7 \nl
\sidehead{12/97}

U(5) &  -1.81/0.06 \ &  -2.46/0.02 \ & 3.06/0.04 &  116.8/\ 0.5 & 0.12/\ 1.4 \nl

B(5) &  -1.93/0.04 \ &  -2.62/0.03 \ & 3.26/0.03 &  116.8/\ 0.4 & 0.06/\ 0.6 \nl

V(5) &  -2.02/0.02 \ &  -2.78/0.05 \ & 3.44/0.04 &  117.0/\ 0.4 & 0.10/\ 0.5 \nl

R(5) &  -1.95/0.04 \ &  -2.67/0.04 \ & 3.31/0.04 &  116.9/\ 0.4 & 0.05/\ 0.7 \nl

I(5) &  -1.74/0.08 \ &  -2.38/0.10 \ & 2.95/0.09 &  116.9/\ 1.0 & 0.08/\ 0.8 \nl

\enddata
\end{deluxetable}

\begin{deluxetable}{lllccc}
\footnotesize
\tt
\tablecolumns{6}
\tablewidth{0pt}
\tablecaption{o Sco}
\tablehead{
\colhead{Date(Mo/Yr)} &
\colhead{$q\/$/$dq$} &
\colhead{$u\/$/$du$} &
\colhead{$p\/$/$dp$} &
\colhead{$\theta\/$/$d\theta$} &
\colhead{$dpi\/$/$d\theta i$} \nl
\colhead{Filter($n$)} &
\colhead{(\%)} &
\colhead{(\%)} &
\colhead{(\%)} &
\colhead{($\arcdeg$)} &
\colhead{(\%) ($\arcdeg$)}
}

\startdata
\sidehead{06/95}

U(4) & \ 1.25/0.28 \ & \ 2.40/0.17 \ & 2.73/0.18 & \ 31.2/\ 2.8 & 0.17/\ 3.2 \nl

B(4) & \ 1.58/0.13 \ & \ 3.03/0.06 \ & 3.42/0.11 & \ 31.3/\ 0.8 & 0.09/\ 1.1 \nl

V(4) & \ 1.96/0.08 \ & \ 3.72/0.10 \ & 4.21/0.11 & \ 31.2/\ 0.4 & 0.11/\ 1.0 \nl

R(4) & \ 2.05/0.05 \ & \ 3.94/0.06 \ & 4.44/0.08 & \ 31.3/\ 0.2 & 0.11/\ 0.7 \nl

I(4) & \ 1.93/0.03 \ & \ 3.85/0.08 \ & 4.31/0.07 & \ 31.7/\ 0.3 & 0.19/\ 1.2 \nl
\sidehead{06/96}

U(4) & \ 0.93/0.12 \ & \ 2.36/0.38 \ & 2.54/0.39 & \ 34.3/\ 0.8 & 0.19/\ 2.1 \nl

B(4) & \ 1.37/0.05 \ & \ 3.14/0.08 \ & 3.43/0.10 & \ 33.2/\ 0.3 & 0.06/\ 0.8 \nl

V(4) & \ 1.68/0.10 \ & \ 3.81/0.05 \ & 4.17/0.01 & \ 33.1/\ 0.8 & 0.10/\ 0.9 \nl

R(4) & \ 1.76/0.08 \ & \ 4.07/0.07 \ & 4.44/0.07 & \ 33.3/\ 0.5 & 0.09/\ 0.8 \nl

I(4) & \ 1.70/0.08 \ & \ 3.98/0.03 \ & 4.33/0.04 & \ 33.4/\ 0.5 & 0.06/\ 0.8 \nl
\sidehead{07/97}

U(5) & \ 1.05/0.16 \ & \ 2.34/0.20 \ & 2.58/0.23 & \ 33.0/\ 1.5 & 0.24/\ 2.5 \nl

B(5) & \ 1.34/0.05 \ & \ 3.14/0.04 \ & 3.41/0.05 & \ 33.5/\ 0.4 & 0.06/\ 0.5 \nl

V(5) & \ 1.57/0.12 \ & \ 3.80/0.10 \ & 4.11/0.12 & \ 33.8/\ 0.7 & 0.09/\ 0.7 \nl

R(5) & \ 1.68/0.10 \ & \ 4.03/0.08 \ & 4.36/0.07 & \ 33.7/\ 0.7 & 0.05/\ 0.5 \nl

I(5) & \ 1.56/0.13 \ & \ 4.06/0.05 \ & 4.35/0.07 & \ 34.5/\ 0.8 & 0.15/\ 0.9 \nl
\sidehead{07/98}

U(4) & \ 1.00/0.15 \ & \ 2.29/0.17 \ & 2.51/0.18 & \ 33.3/\ 1.7 & 0.15/\ 3.1 \nl

B(4) & \ 1.37/0.07 \ & \ 3.06/0.05 \ & 3.35/0.06 & \ 33.0/\ 0.5 & 0.09/\ 0.6 \nl

V(4) & \ 1.72/0.02 \ & \ 3.73/0.08 \ & 4.11/0.08 & \ 32.6/\ 0.2 & 0.09/\ 0.6 \nl

R(4) & \ 1.82/0.07 \ & \ 3.97/0.07 \ & 4.37/0.08 & \ 32.7/\ 0.4 & 0.11/\ 0.6 \nl

I(4) & \ 1.75/0.06 \ & \ 3.85/0.03 \ & 4.23/0.03 & \ 32.8/\ 0.4 & 0.15/\ 1.0 \nl

\enddata
\end{deluxetable}

\begin{deluxetable}{lllccc}
\footnotesize
\tt
\tablecolumns{6}
\tablewidth{0pt}
\tablecaption{$\gamma$ Cas}
\tablehead{
\colhead{Date(Mo/Yr)} &
\colhead{$q\/$/$dq$} &
\colhead{$u\/$/$du$} &
\colhead{$p\/$/$dp$} &
\colhead{$\theta\/$/$d\theta$} &
\colhead{$dpi\/$/$d\theta i$} \nl
\colhead{Filter($n$)} &
\colhead{(\%)} &
\colhead{(\%)} &
\colhead{(\%)} &
\colhead{($\arcdeg$)} &
\colhead{(\%) ($\arcdeg$)}
}

\startdata
\sidehead{12/94}

U(4) & -0.43/0.08 \ & -0.24/0.08 \ & 0.50/0.10 & 103.9/\ 3.1 & 0.07/\ 2.5 \nl

B(4) & -0.66/0.05 \ & -0.42/0.02 \ & 0.79/0.04 & 106.1/\ 1.4 & 0.05/\ 1.7 \nl

V(4) & -0.58/0.04 \ & -0.35/0.03 \ & 0.69/0.02 & 105.7/\ 2.1 & 0.05/\ 1.8 \nl

R(4) & -0.62/0.06 \ & -0.32/0.03 \ & 0.70/0.05 & 103.8/\ 2.2 & 0.05/\ 1.5 \nl

I(4) & -0.54/0.05 \ & -0.22/0.04 \ & 0.59/0.05 & 100.8/\ 1.4 & 0.06/\ 3.6 \nl
\sidehead{01/96}

U(3) & -0.56/0.08 \ & -0.28/0.04 \ & 0.63/0.08 & 103.5/\ 1.8 & 0.08/\ 2.1 \nl

B(3) & -0.70/0.05 \ & -0.50/0.12 \ & 0.87/0.06 & 107.7/\ 3.9 & 0.03/\ 2.5 \nl

V(3) & -0.59/0.03 \ & -0.40/0.05 \ & 0.72/0.01 & 107.0/\ 2.2 & 0.05/\ 1.5 \nl

R(3) & -0.60/0.08 \ & -0.31/0.09 \ & 0.68/0.08 & 103.7/\ 3.8 & 0.06/\ 2.1 \nl

I(3) & -0.41/0.04 \ & -0.25/0.03 \ & 0.49/0.05 & 106.4/\ 1.4 & 0.09/\ 3.5 \nl
\sidehead{12/96}

U(5) & -0.38/0.05 \ & -0.30/0.08 \ & 0.49/0.07 & 109.3/\ 3.1 & 0.07/\ 3.7 \nl

B(5) & -0.61/0.06 \ & -0.42/0.08 \ & 0.74/0.06 & 107.3/\ 3.1 & 0.07/\ 2.9 \nl

V(5) & -0.56/0.08 \ & -0.33/0.09 \ & 0.66/0.09 & 105.3/\ 3.1 & 0.06/\ 2.6 \nl

R(5) & -0.55/0.08 \ & -0.28/0.08 \ & 0.62/0.07 & 103.2/\ 3.8 & 0.10/\ 3.2 \nl

I(5) & -0.46/0.06 \ & -0.24/0.06 \ & 0.52/0.07 & 103.2/\ 3.0 & 0.09/\ 3.2 \nl
\sidehead{12/97}

U(6) & -0.50/0.08 \ & -0.31/0.03 \ & 0.60/0.08 & 106.2/\ 1.6 & 0.08/\ 4.9 \nl

B(6) & -0.64/0.04 \ & -0.39/0.09 \ & 0.76/0.08 & 105.6/\ 2.5 & 0.11/\ 2.9 \nl

V(6) & -0.58/0.06 \ & -0.40/0.06 \ & 0.70/0.08 & 107.2/\ 1.2 & 0.08/\ 2.3 \nl

R(6) & -0.58/0.06 \ & -0.27/0.08 \ & 0.65/0.04 & 102.4/\ 4.0 & 0.11/\ 3.4 \nl

I(6) & -0.45/0.05 \ & -0.20/0.05 \ & 0.50/0.05 & 102.2/\ 2.6 & 0.06/\ 6.8 \nl

\enddata
\end{deluxetable}

\begin{deluxetable}{lllccc}
\footnotesize
\tt
\tablecolumns{6}
\tablewidth{0pt}
\tablecaption{$\phi$ Per}
\tablehead{
\colhead{Date(Mo/Yr)} &
\colhead{$q\/$/$dq$} &
\colhead{$u\/$/$du$} &
\colhead{$p\/$/$dp$} &
\colhead{$\theta\/$/$d\theta$} &
\colhead{$dpi\/$/$d\theta i$} \nl
\colhead{Filter($n$)} &
\colhead{(\%)} &
\colhead{(\%)} &
\colhead{(\%)} &
\colhead{($\arcdeg$)} &
\colhead{(\%) ($\arcdeg$)}
}

\startdata
\sidehead{12/94}

U(4) & \ 0.25/0.06 \ & \ 0.85/0.10 \ & 0.89/0.10 & \ 36.9/\ 2.2 & 0.08/\ 2.0 \nl

B(4) & \ 0.46/0.11 \ & \ 1.09/0.08 \ & 1.19/0.08 & \ 33.6/\ 2.5 & 0.04/\ 1.7 \nl

V(4) & \ 0.34/0.08 \ & \ 0.99/0.08 \ & 1.05/0.09 & \ 35.5/\ 1.8 & 0.05/\ 1.5 \nl

R(4) & \ 0.24/0.07 \ & \ 0.90/0.08 \ & 0.94/0.07 & \ 37.5/\ 2.2 & 0.06/\ 2.6 \nl

I(4) & \ 0.10/0.08 \ & \ 0.72/0.03 \ & 0.73/0.03 & \ 41.0/\ 3.0 & 0.10/\ 2.4 \nl \sidehead{01/96}

U(3) & \ 0.04/0.08 \ & \ 0.85/0.11 \ & 0.86/0.11 & \ 43.7/\ 2.5 & 0.07/\ 3.8 \nl

B(3) & \ 0.41/0.06 \ & \ 1.23/0.06 \ & 1.30/0.07 & \ 35.8/\ 1.2 & 0.06/\ 1.2 \nl

V(3) & \ 0.37/0.07 \ & \ 1.10/0.08 \ & 1.16/0.09 & \ 35.8/\ 1.4 & 0.07/\ 1.1 \nl

R(3) & \ 0.25/0.03 \ & \ 0.94/0.09 \ & 0.97/0.08 & \ 37.5/\ 1.2 & 0.06/\ 1.4 \nl

I(3) & \ 0.22/0.03 \ & \ 0.69/0.08 \ & 0.73/0.09 & \ 36.5/\ 0.4 & 0.11/\ 2.6 \nl
\sidehead{12/96}

U(5) & \ 0.18/0.05 \ & \ 0.81/0.12 \ & 0.83/0.12 & \ 38.5/\ 1.6 & 0.10/\ 2.1 \nl

B(5) & \ 0.50/0.04 \ & \ 1.28/0.10 \ & 1.37/0.08 & \ 34.3/\ 1.5 & 0.05/\ 1.4 \nl

V(5) & \ 0.39/0.06 \ & \ 1.12/0.10 \ & 1.19/0.10 & \ 35.4/\ 1.3 & 0.07/\ 1.6 \nl

R(5) & \ 0.25/0.05 \ & \ 1.01/0.10 \ & 1.05/0.10 & \ 37.8/\ 1.5 & 0.05/\ 2.6 \nl

I(5) & \ 0.19/0.07 \ & \ 0.81/0.17 \ & 0.84/0.17 & \ 38.1/\ 2.3 & 0.06/\ 3.2 \nl
\sidehead{12/97}

U(6) & \ 0.15/0.07 \ & \ 0.83/0.08 \ & 0.85/0.08 & \ 39.7/\ 2.1 & 0.07/\ 3.8 \nl

B(6) & \ 0.52/0.01 \ & \ 1.34/0.06 \ & 1.44/0.06 & \ 34.4/\ 0.4 & 0.05/\ 1.3 \nl

V(6) & \ 0.44/0.04 \ & \ 1.21/0.04 \ & 1.29/0.05 & \ 35.0/\ 0.8 & 0.07/\ 2.3 \nl

R(6) & \ 0.33/0.02 \ & \ 1.09/0.08 \ & 1.14/0.07 & \ 36.6/\ 0.7 & 0.05/\ 1.8 \nl

I(6) & \ 0.29/0.05 \ & \ 0.95/0.08 \ & 1.00/0.08 & \ 36.5/\ 1.7 & 0.09/\ 3.2 \nl

\enddata
\end{deluxetable}

\begin{deluxetable}{lllccc}
\footnotesize
\tt
\tablecolumns{6}
\tablewidth{0pt}
\tablecaption{48 Per}
\tablehead{
\colhead{Date(Mo/Yr)} &
\colhead{$q\/$/$dq$} &
\colhead{$u\/$/$du$} &
\colhead{$p\/$/$dp$} &
\colhead{$\theta\/$/$d\theta$} &
\colhead{$dpi\/$/$d\theta i$} \nl
\colhead{Filter($n$)} &
\colhead{(\%)} &
\colhead{(\%)} &
\colhead{(\%)} &
\colhead{($\arcdeg$)} &
\colhead{(\%) ($\arcdeg$)}
}

\startdata
\sidehead{12/94}

U(4) & \ 0.77/0.05 \ & -0.21/0.06 \ & 0.81/0.04 & 172.3/\ 2.7 & 0.11/\ 2.2 \nl

B(4) & \ 0.76/0.06 \ & -0.28/0.05 \ & 0.81/0.04 & 169.7/\ 2.3 & 0.05/\ 1.8 \nl

V(4) & \ 0.91/0.07 \ & -0.34/0.07 \ & 0.97/0.05 & 169.9/\ 2.4 & 0.09/\ 1.8 \nl

R(4) & \ 0.85/0.05 \ & -0.29/0.03 \ & 0.90/0.05 & 170.6/\ 1.2 & 0.07/\ 1.3 \nl

I(4) & \ 0.72/0.05 \ & -0.23/0.06 \ & 0.76/0.05 & 171.2/\ 2.2 & 0.06/\ 2.8 \nl \sidehead{01/96}

U(3) & \ 0.61/0.06 \ & -0.14/0.04 \ & 0.63/0.07 & 173.3/\ 2.1 & 0.09/\ 3.3 \nl

B(3) & \ 0.82/0.03 \ & -0.20/0.06 \ & 0.85/0.03 & 173.2/\ 1.8 & 0.03/\ 1.8 \nl

V(3) & \ 0.89/0.05 \ & -0.20/0.06 \ & 0.91/0.04 & 173.6/\ 2.1 & 0.05/\ 3.0 \nl

R(3) & \ 0.92/0.05 \ & -0.24/0.04 \ & 0.96/0.05 & 172.6/\ 1.4 & 0.07/\ 0.8 \nl

I(3) & \ 0.81/0.04 \ & -0.17/0.05 \ & 0.83/0.04 & 174.1/\ 1.8 & 0.08/\ 1.3 \nl
\sidehead{12/96}

U(4) & \ 0.83/0.07 \ & -0.24/0.08 \ & 0.87/0.05 & 171.6/\ 3.3 & 0.08/\ 2.0 \nl

B(4) & \ 0.90/0.02 \ & -0.25/0.04 \ & 0.94/0.02 & 172.1/\ 1.2 & 0.07/\ 1.4 \nl

V(4) & \ 0.96/0.07 \ & -0.24/0.06 \ & 0.99/0.08 & 172.9/\ 1.7 & 0.06/\ 2.4 \nl

R(4) & \ 0.97/0.04 \ & -0.21/0.09 \ & 1.00/0.02 & 173.9/\ 2.7 & 0.07/\ 2.0 \nl

I(4) & \ 0.85/0.09 \ & -0.25/0.08 \ & 0.90/0.08 & 171.5/\ 2.9 & 0.11/\ 4.4 \nl
\sidehead{12/97}

U(6) & \ 0.71/0.06 \ & -0.23/0.07 \ & 0.75/0.07 & 170.9/\ 2.0 & 0.08/\ 2.9 \nl

B(6) & \ 0.83/0.04 \ & -0.22/0.05 \ & 0.86/0.04 & 172.5/\ 1.3 & 0.06/\ 2.2 \nl

V(6) & \ 0.97/0.08 \ & -0.23/0.03 \ & 1.00/0.08 & 173.3/\ 0.8 & 0.08/\ 2.0 \nl

R(6) & \ 0.90/0.03 \ & -0.21/0.06 \ & 0.93/0.03 & 173.5/\ 2.0 & 0.08/\ 2.3 \nl

I(6) & \ 0.88/0.12 \ & -0.17/0.07 \ & 0.90/0.12 & 174.4/\ 2.6 & 0.10/\ 2.7 \nl

\enddata
\end{deluxetable}

\begin{deluxetable}{lllccc}
\footnotesize
\tt
\tablecolumns{6}
\tablewidth{0pt}
\tablecaption{$\zeta$ Tau}
\tablehead{
\colhead{Date(Mo/Yr)} &
\colhead{$q\/$/$dq$} &
\colhead{$u\/$/$du$} &
\colhead{$p\/$/$dp$} &
\colhead{$\theta\/$/$d\theta$} &
\colhead{$dpi\/$/$d\theta i$} \nl
\colhead{Filter($n$)} &
\colhead{(\%)} &
\colhead{(\%)} &
\colhead{(\%)} &
\colhead{($\arcdeg$)} &
\colhead{(\%) ($\arcdeg$)}
}

\startdata
\sidehead{12/94}

U(4) & \ 0.51/0.06 \ & \ 1.12/0.06 \ & 1.24/0.06 & \ 32.8/\ 1.3 & 0.07/\ 1.2 \nl

B(4) & \ 0.57/0.08 \ & \ 1.31/0.08 \ & 1.43/0.10 & \ 33.3/\ 1.0 & 0.04/\ 0.9 \nl

V(4) & \ 0.63/0.06 \ & \ 1.24/0.05 \ & 1.39/0.04 & \ 31.5/\ 1.6 & 0.03/\ 0.7 \nl

R(4) & \ 0.49/0.01 \ & \ 1.15/0.02 \ & 1.26/0.02 & \ 33.4/\ 0.3 & 0.06/\ 1.3 \nl

I(4) & \ 0.31/0.06 \ & \ 1.05/0.10 \ & 1.10/0.10 & \ 36.7/\ 1.6 & 0.07/\ 1.4 \nl \sidehead{01/96}

U(3) & \ 0.44/0.04 \ & \ 1.01/0.08 \ & 1.10/0.06 & \ 33.2/\ 1.5 & 0.04/\ 1.2 \nl

B(3) & \ 0.65/0.01 \ & \ 1.26/0.10 \ & 1.42/0.08 & \ 31.3/\ 1.0 & 0.05/\ 1.3 \nl

V(3) & \ 0.59/0.05 \ & \ 1.25/0.08 \ & 1.38/0.08 & \ 32.4/\ 0.3 & 0.05/\ 1.7 \nl

R(3) & \ 0.51/0.02 \ & \ 1.20/0.02 \ & 1.30/0.02 & \ 33.5/\ 0.4 & 0.10/\ 1.5 \nl

I(3) & \ 0.52/0.03 \ & \ 1.06/0.10 \ & 1.18/0.10 & \ 31.9/\ 0.5 & 0.06/\ 1.6 \nl
\sidehead{12/96}

U(4) & \ 0.59/0.02 \ & \ 1.04/0.04 \ & 1.21/0.03 & \ 30.2/\ 0.6 & 0.08/\ 2.9 \nl

B(4) & \ 0.64/0.04 \ & \ 1.42/0.05 \ & 1.57/0.04 & \ 32.8/\ 0.8 & 0.05/\ 2.8 \nl

V(4) & \ 0.59/0.03 \ & \ 1.30/0.05 \ & 1.44/0.03 & \ 32.8/\ 0.9 & 0.06/\ 2.3 \nl

R(4) & \ 0.52/0.04 \ & \ 1.26/0.05 \ & 1.36/0.04 & \ 33.8/\ 1.1 & 0.11/\ 1.5 \nl

I(4) & \ 0.44/0.06 \ & \ 1.01/0.10 \ & 1.11/0.11 & \ 33.3/\ 0.9 & 0.11/\ 2.5 \nl
\sidehead{12/97}

U(6) & \ 0.50/0.08 \ & \ 1.08/0.08 \ & 1.20/0.06 & \ 32.6/\ 2.3 & 0.09/\ 2.5 \nl

B(6) & \ 0.74/0.07 \ & \ 1.45/0.11 \ & 1.63/0.09 & \ 31.5/\ 1.4 & 0.08/\ 2.0 \nl

V(6) & \ 0.71/0.05 \ & \ 1.37/0.12 \ & 1.55/0.10 & \ 31.3/\ 1.5 & 0.10/\ 1.8 \nl

R(6) & \ 0.55/0.07 \ & \ 1.30/0.11 \ & 1.42/0.10 & \ 33.5/\ 1.7 & 0.07/\ 2.0 \nl

I(6) & \ 0.57/0.02 \ & \ 1.12/0.12 \ & 1.26/0.11 & \ 31.4/\ 1.1 & 0.07/\ 2.5 \nl

\enddata
\end{deluxetable}

\begin{deluxetable}{lllccc}
\footnotesize
\tt
\tablecolumns{6}
\tablewidth{0pt}
\tablecaption{48 Lib}
\tablehead{
\colhead{Date(Mo/Yr)} &
\colhead{$q\/$/$dq$} &
\colhead{$u\/$/$du$} &
\colhead{$p\/$/$dp$} &
\colhead{$\theta\/$/$d\theta$} &
\colhead{$dpi\/$/$d\theta i$} \nl
\colhead{Filter($n$)} &
\colhead{(\%)} &
\colhead{(\%)} &
\colhead{(\%)} &
\colhead{($\arcdeg$)} &
\colhead{(\%) ($\arcdeg$)}
}

\startdata
\sidehead{06/95}

U(4) &  -0.40/0.12 \ &  -0.61/0.07 \ & 0.74/0.10 &  118.3/\ 4.4 & 0.13/\ 5.9 \nl

B(4) &  -0.36/0.06 \ &  -0.83/0.06 \ & 0.91/0.05 &  123.3/\ 2.2 & 0.03/\ 2.9 \nl

V(4) &  -0.46/0.03 \ &  -0.71/0.04 \ & 0.85/0.04 &  118.7/\ 1.4 & 0.06/\ 2.2 \nl

R(4) &  -0.43/0.05 \ &  -0.60/0.04 \ & 0.74/0.02 &  117.1/\ 2.3 & 0.07/\ 2.1 \nl

I(4) &  -0.51/0.06 \ &  -0.54/0.13 \ & 0.76/0.10 &  112.6/\ 4.1 & 0.10/\ 4.9 \nl
\sidehead{06/96}

U(4) &  -0.43/0.04 \ &  -0.69/0.09 \ & 0.82/0.06 &  118.9/\ 2.8 & 0.12/\ 3.0 \nl

B(4) &  -0.40/0.03 \ &  -0.87/0.06 \ & 0.96/0.07 &  122.6/\ 0.7 & 0.07/\ 2.6 \nl

V(4) &  -0.52/0.05 \ &  -0.73/0.08 \ & 0.90/0.09 &  117.2/\ 1.0 & 0.06/\ 1.9 \nl

R(4) &  -0.55/0.07 \ &  -0.64/0.06 \ & 0.86/0.07 &  114.7/\ 2.1 & 0.06/\ 2.6 \nl

I(4) &  -0.52/0.02 \ &  -0.51/0.11 \ & 0.74/0.08 &  112.2/\ 3.1 & 0.12/\ 5.4 \nl
\sidehead{07/97}

U(4) &  -0.35/0.10 \ &  -0.41/0.06 \ & 0.56/0.06 &  114.8/\ 5.0 & 0.12/\ 6.0 \nl

B(4) &  -0.46/0.05 \ &  -0.72/0.08 \ & 0.86/0.08 &  118.5/\ 1.5 & 0.09/\ 1.6 \nl

V(4) &  -0.60/0.04 \ &  -0.65/0.05 \ & 0.88/0.06 &  113.8/\ 1.2 & 0.11/\ 1.9 \nl

R(4) &  -0.57/0.09 \ &  -0.63/0.05 \ & 0.85/0.09 &  114.0/\ 1.4 & 0.09/\ 2.6 \nl

I(4) &  -0.67/0.11 \ &  -0.60/0.09 \ & 0.91/0.12 &  111.4/\ 2.1 & 0.11/\ 5.3 \nl
\sidehead{07/98}

U(5) &  -0.35/0.08 \ &  -0.53/0.08 \ & 0.65/0.10 &  118.5/\ 2.3 & 0.09/\ 5.4 \nl

B(5) &  -0.42/0.06 \ &  -0.78/0.08 \ & 0.89/0.06 &  120.7/\ 2.6 & 0.04/\ 3.9 \nl

V(5) &  -0.51/0.04 \ &  -0.71/0.08 \ & 0.88/0.08 &  117.1/\ 1.0 & 0.07/\ 3.0 \nl

R(5) &  -0.53/0.02 \ &  -0.63/0.06 \ & 0.83/0.04 &  114.8/\ 1.8 & 0.09/\ 2.3 \nl

I(5) &  -0.42/0.03 \ &  -0.58/0.09 \ & 0.74/0.06 &  117.2/\ 3.0 & 0.14/\ 6.6 \nl

\enddata
\end{deluxetable}

\begin{deluxetable}{lllccc}
\footnotesize
\tt
\tablecolumns{6}
\tablewidth{0pt}
\tablecaption{$\chi$ Oph}
\tablehead{
\colhead{Date(Mo/Yr)} &
\colhead{$q\/$/$dq$} &
\colhead{$u\/$/$du$} &
\colhead{$p\/$/$dp$} &
\colhead{$\theta\/$/$d\theta$} &
\colhead{$dpi\/$/$d\theta i$} \nl
\colhead{Filter($n$)} &
\colhead{(\%)} &
\colhead{(\%)} &
\colhead{(\%)} &
\colhead{($\arcdeg$)} &
\colhead{(\%) ($\arcdeg$)}
}

\startdata
\sidehead{06/95}

U(4) & \ 0.03/0.06 \ &  -0.39/0.05 \ & 0.41/0.06 &  137.6/\ 4.9 & 0.11/\ 8.5 \nl

B(4) &  -0.06/0.06 \ &  -0.52/0.08 \ & 0.54/0.08 &  132.1/\ 2.9 & 0.06/\ 5.0 \nl

V(4) & \ 0.00/0.05 \ &  -0.52/0.06 \ & 0.53/0.06 &  134.8/\ 2.9 & 0.10/\ 5.4 \nl

R(4) &  -0.04/0.07 \ &  -0.46/0.07 \ & 0.47/0.07 &  132.7/\ 4.4 & 0.05/\ 4.8 \nl

I(4) &  -0.02/0.04 \ &  -0.43/0.06 \ & 0.45/0.04 &  136.0/\ 3.5 & 0.15/12.1 \nl 
\sidehead{06/96}

U(4) &  -0.03/0.20 \ &  -0.28/0.03 \ & 0.38/0.05 &  134.9/18.7 & 0.12/15.7 \nl

B(4) & \ 0.00/0.08 \ &  -0.42/0.08 \ & 0.44/0.07 &  135.5/\ 8.3 & 0.09/\ 9.2 \nl

V(4) & \ 0.01/0.08 \ &  -0.45/0.07 \ & 0.46/0.07 &  135.8/\ 4.9 & 0.13/\ 3.9 \nl

R(4) &  -0.05/0.07 \ &  -0.47/0.06 \ & 0.48/0.06 &  131.8/\ 3.7 & 0.14/\ 6.0 \nl

I(4) &  -0.16/0.06 \ &  -0.41/0.05 \ & 0.47/0.05 &  126.0/\ 5.7 & 0.10/12.4 \nl
\sidehead{07/97}

U(4) & \ 0.12/0.11 \ &  -0.28/0.05 \ & 0.36/0.05 &  144.7/\ 9.8 & 0.08/18.9 \nl

B(4) &  -0.01/0.12 \ &  -0.45/0.05 \ & 0.47/0.05 &  135.0/\ 7.4 & 0.08/\ 4.6 \nl

V(4) & \ 0.00/0.19 \ &  -0.44/0.07 \ & 0.48/0.07 &  136.2/11.5 & 0.08/\ 5.5 \nl

R(4) & \ 0.05/0.13 \ &  -0.48/0.07 \ & 0.50/0.06 &  138.7/\ 7.7 & 0.07/\ 7.6 \nl

I(4) &  -0.04/0.11 \ &  -0.38/0.11 \ & 0.41/0.10 &  134.5/10.6 & 0.13/\ 6.1 \nl
\sidehead{07/98}

U(5) & \ 0.03/0.13 \ &  -0.37/0.08 \ & 0.40/0.10 &  134.7/\ 7.6 & 0.13/10.7 \nl

B(5) & \ 0.01/0.07 \ &  -0.47/0.05 \ & 0.48/0.05 &  135.6/\ 4.6 & 0.12/\ 4.8 \nl

V(5) &  -0.03/0.04 \ &  -0.47/0.06 \ & 0.49/0.06 &  133.4/\ 2.7 & 0.10/\ 7.9 \nl

R(5) & \ 0.01/0.07 \ &  -0.47/0.03 \ & 0.48/0.03 &  135.6/\ 3.6 & 0.08/\ 7.1 \nl

I(5) & \ 0.01/0.14 \ &  -0.46/0.08 \ & 0.50/0.07 &  135.5/\ 8.4 & 0.12/\ 8.6 \nl

\enddata
\end{deluxetable}

\begin{deluxetable}{lllccc}
\footnotesize
\tt
\tablecolumns{6}
\tablewidth{0pt}
\tablecaption{$\pi$ Aqr}
\tablehead{
\colhead{Date(Mo/Yr)} &
\colhead{$q\/$/$dq$} &
\colhead{$u\/$/$du$} &
\colhead{$p\/$/$dp$} &
\colhead{$\theta\/$/$d\theta$} &
\colhead{$dpi\/$/$d\theta i$} \nl
\colhead{Filter($n$)} &
\colhead{(\%)} &
\colhead{(\%)} &
\colhead{(\%)} &
\colhead{($\arcdeg$)} &
\colhead{(\%) ($\arcdeg$)}
}

\startdata
\sidehead{06/95}

U(3) &  -0.01/0.11 \ &  -0.44/0.15 \ & 0.47/0.14 &  131.6/\ 7.0 & 0.13/12.1 \nl

B(3) &  -0.08/0.07 \ &  -0.50/0.04 \ & 0.51/0.04 &  130.0/\ 3.8 & 0.10/\ 5.1 \nl

V(3) &  -0.06/0.05 \ &  -0.48/0.00 \ & 0.49/0.01 &  130.9/\ 3.0 & 0.09/\ 4.7 \nl

R(3) &  -0.05/0.04 \ &  -0.45/0.04 \ & 0.46/0.03 &  132.0/\ 2.5 & 0.09/\ 4.0 \nl

I(3) & \ 0.04/0.03 \ &  -0.40/0.01 \ & 0.42/0.02 &  137.8/\ 2.4 & 0.08/\ 9.4 \nl \sidehead{06/96}

U(4) &  -0.34/0.03 \ &  -0.32/0.05 \ & 0.48/0.04 &  112.2/\ 2.3 & 0.10/\ 5.2 \nl

B(4) &  -0.38/0.04 \ &  -0.31/0.02 \ & 0.50/0.02 &  109.6/\ 2.3 & 0.07/\ 5.4 \nl

V(4) &  -0.41/0.11 \ &  -0.30/0.07 \ & 0.52/0.06 &  108.9/\ 6.0 & 0.08/\ 3.6 \nl

R(4) &  -0.37/0.08 \ &  -0.29/0.03 \ & 0.48/0.06 &  109.5/\ 3.8 & 0.09/\ 4.6 \nl

I(4) &  -0.29/0.14 \ &  -0.31/0.04 \ & 0.46/0.08 &  114.2/\ 9.5 & 0.14/10.4 \nl \sidehead{07/97}

U(4) &  -0.26/0.02 \ &  -0.29/0.05 \ & 0.40/0.05 &  113.0/\ 2.8 & 0.10/\ 7.8 \nl

B(4) &  -0.36/0.02 \ &  -0.34/0.07 \ & 0.50/0.06 &  111.8/\ 2.5 & 0.08/\ 3.1 \nl

V(4) &  -0.40/0.04 \ &  -0.31/0.03 \ & 0.51/0.03 &  108.8/\ 2.4 & 0.04/\ 2.8 \nl

R(4) &  -0.37/0.06 \ &  -0.31/0.03 \ & 0.49/0.03 &  110.1/\ 3.2 & 0.08/\ 3.4 \nl

I(4) &  -0.31/0.08 \ &  -0.26/0.08 \ & 0.43/0.07 &  110.4/\ 7.0 & 0.07/\ 9.6 \nl
 \sidehead{07/98}

U(5) &  -0.37/0.10 \ &  -0.27/0.09 \ & 0.48/0.11 &  107.6/\ 4.9 & 0.10/\ 6.4 \nl

B(5) &  -0.36/0.06 \ &  -0.32/0.03 \ & 0.49/0.05 &  110.8/\ 2.5 & 0.05/\ 2.4 \nl

V(5) &  -0.40/0.04 \ &  -0.30/0.05 \ & 0.51/0.02 &  108.6/\ 3.5 & 0.08/\ 4.5 \nl

R(5) &  -0.41/0.06 \ &  -0.27/0.07 \ & 0.50/0.08 &  106.7/\ 2.4 & 0.07/\ 4.5 \nl

I(5) &  -0.33/0.11 \ &  -0.25/0.08 \ & 0.44/0.10 &  110.8/\ 5.6 & 0.09/\ 9.5 \nl

\enddata
\end{deluxetable}

\begin{deluxetable}{lllccc}
\footnotesize
\tt
\tablecolumns{6}
\tablewidth{0pt}
\tablecaption{o And}
\tablehead{
\colhead{Date(Mo/Yr)} &
\colhead{$q\/$/$dq$} &
\colhead{$u\/$/$du$} &
\colhead{$p\/$/$dp$} &
\colhead{$\theta\/$/$d\theta$} &
\colhead{$dpi\/$/$d\theta i$} \nl
\colhead{Filter($n$)} &
\colhead{(\%)} &
\colhead{(\%)} &
\colhead{(\%)} &
\colhead{($\arcdeg$)} &
\colhead{(\%) ($\arcdeg$)}
}

\startdata
\sidehead{06/95}

U(3) &  -0.48/0.02 \ &  -0.11/0.03 \ & 0.50/0.01 & \ 96.5/\ 1.8 & 0.07/\ 5.0 \nl

B(3) &  -0.56/0.07 \ &  -0.16/0.11 \ & 0.59/0.09 & \ 97.8/\ 4.4 & 0.03/\ 3.2 \nl

V(3) &  -0.53/0.01 \ &  -0.05/0.01 \ & 0.54/0.02 & \ 92.6/\ 0.7 & 0.06/\ 1.0 \nl

R(3) &  -0.48/0.03 \ &  -0.03/0.02 \ & 0.49/0.03 & \ 91.8/\ 1.0 & 0.04/\ 2.9 \nl

I(3) &  -0.46/0.09 \ &  -0.06/0.07 \ & 0.48/0.10 & \ 93.6/\ 3.4 & 0.05/\ 7.6 \nl \sidehead{06/96}

U(4) &  -0.40/0.07 \ &  -0.06/0.07 \ & 0.41/0.07 & \ 93.8/\ 4.7 & 0.09/\ 6.6 \nl

B(4) &  -0.37/0.04 \ &  -0.05/0.06 \ & 0.38/0.04 & \ 93.6/\ 3.6 & 0.08/\ 3.6 \nl

V(4) &  -0.32/0.06 \ &  -0.07/0.07 \ & 0.35/0.05 & \ 96.7/\ 5.7 & 0.05/\ 9.2 \nl

R(4) &  -0.29/0.07 \ &  -0.06/0.04 \ & 0.31/0.06 & \ 95.6/\ 4.0 & 0.07/\ 9.8 \nl

I(4) &  -0.30/0.10 \ &  -0.04/0.08 \ & 0.32/0.11 & \ 94.3/\ 6.0 & 0.11/\ 8.9 \nl
\sidehead{07/97}

U(4) &  -0.27/0.03 \ & \ 0.00/0.06 \ & 0.28/0.03 & \ 89.7/\ 6.4 & 0.10/\ 7.5 \nl

B(4) &  -0.36/0.02 \ &  -0.05/0.05 \ & 0.37/0.02 & \ 94.1/\ 4.6 & 0.06/\ 4.4 \nl

V(4) &  -0.36/0.05 \ &  -0.02/0.01 \ & 0.37/0.05 & \ 91.0/\ 1.1 & 0.06/\ 7.4 \nl

R(4) &  -0.35/0.04 \ &  -0.04/0.02 \ & 0.36/0.04 & \ 93.1/\ 1.7 & 0.04/\ 5.3 \nl

I(4) &  -0.30/0.08 \ & \ 0.04/0.08 \ & 0.32/0.08 & \ 88.0/\ 7.8 & 0.05/\ 9.5 \nl
\sidehead{07/98}

U(5) &  -0.43/0.06 \ &  -0.26/0.05 \ & 0.51/0.06 &  105.4/\ 2.9 & 0.08/\ 3.6 \nl

B(5) &  -0.57/0.04 \ &  -0.27/0.04 \ & 0.64/0.04 &  102.9/\ 1.9 & 0.07/\ 4.2 \nl

V(5) &  -0.60/0.06 \ &  -0.19/0.08 \ & 0.64/0.07 & \ 98.8/\ 3.0 & 0.07/\ 4.2 \nl

R(5) &  -0.59/0.05 \ &  -0.19/0.07 \ & 0.62/0.06 & \ 98.9/\ 2.8 & 0.07/\ 3.3 \nl

I(5) &  -0.56/0.07 \ &  -0.18/0.06 \ & 0.60/0.07 & \ 99.2/\ 3.1 & 0.13/\ 4.9 \nl

\enddata
\end{deluxetable}

\begin{deluxetable}{lllllcccccc}
\scriptsize
\tt
\tablecolumns{11}
\tablewidth{0pt}
\tablecaption{Winter Be Star Summary}
\tablehead{
\colhead{Star} &
\colhead{$q\/$/$dq$} &
\colhead{<$dq$>} &
\colhead{$u\/$/$du$} &
\colhead{<$du$>} &
\colhead{$p\/$/$dp$} &
\colhead{<$dp$>} &
\colhead{$\theta\/$/$d\theta$} &
\colhead{<$d\theta$>} &
\colhead{<$dpi$>} &
\colhead{<$d\theta i$>} \nl
\colhead{Filter} &
\colhead{(\%)} &
\colhead{(\%)} &
\colhead{(\%)} &
\colhead{(\%)} &
\colhead{(\%)} &
\colhead{(\%)} &
\colhead{($\arcdeg$)} & 
\colhead{($\arcdeg$)} &
\colhead{(\%)} &
\colhead{($\arcdeg$)}
}

\startdata
\sidehead{$\gamma$ Cas}

U & -0.47/0.08 & 0.07 & -0.28/0.03 & 0.06 & 0.56/0.07 & 0.08 & 105.7/ 2.7 &  2.4 & 0.08 &  3.3 \nl

B & -0.65/0.04 & 0.05 & -0.43/0.05 & 0.08 & 0.79/0.06 & 0.06 & 106.7/ 1.0 &  2.7 & 0.06 &  2.5 \nl

V & -0.58/0.01 & 0.05 & -0.37/0.04 & 0.06 & 0.69/0.02 & 0.05 & 106.3/ 0.9 &  2.2 & 0.06 &  2.0 \nl

R & -0.59/0.03 & 0.07 & -0.29/0.02 & 0.07 & 0.66/0.03 & 0.06 & 103.3/ 0.6 &  3.5 & 0.08 &  2.6 \nl

I & -0.47/0.05 & 0.05 & -0.23/0.02 & 0.04 & 0.52/0.05 & 0.05 & 103.2/ 2.4 &  2.1 & 0.08 &  4.3 \nl

GAV & \ \ \ \ \ \ 0.04 \ & 0.06 \ & \ \ \ \ \ \ 0.03 \ & 0.06 \ &
\multicolumn{1}{r}{0.05} & 0.06 & \multicolumn{1}{r}{1.5} &  2.6 & 0.07 &  2.9 \nl

\sidehead{$\phi$ Per}

U & \ 0.16/0.09 & 0.06 & \ 0.83/0.02 & 0.10 & 0.86/0.02 & 0.10 & \ 39.7/ 2.9 &  2.1 & 0.08 &  2.9 \nl

B & \ 0.47/0.05 & 0.06 & \ 1.24/0.11 & 0.08 & 1.33/0.11 & 0.07 & \ 34.5/ 0.9 &  1.4 & 0.05 &  1.4 \nl

V & \ 0.38/0.04 & 0.06 & \ 1.11/0.09 & 0.07 & 1.17/0.10 & 0.08 & \ 35.4/ 0.3 &  1.3 & 0.06 &  1.6 \nl

R & \ 0.27/0.04 & 0.04 & \ 0.99/0.08 & 0.09 & 1.02/0.09 & 0.08 & \ 37.3/ 0.5 &  1.4 & 0.05 &  2.1 \nl

I & \ 0.20/0.08 & 0.06 & \ 0.79/0.12 & 0.09 & 0.82/0.13 & 0.09 & \ 38.0/ 2.1 &  1.8 & 0.09 &  2.8 \nl

GAV & \ \ \ \ \ \ 0.06 \ & 0.06 \ & \ \ \ \ \ \ 0.08 \ & 0.09 \ &
\multicolumn{1}{r}{0.09} & 0.09 & \multicolumn{1}{r}{1.4} &  1.6 & 0.07 &  2.2 \nl

\sidehead{48 Per}

U & \ 0.73/0.09 & 0.06 & -0.20/0.05 & 0.06 & 0.76/0.10 & 0.06 & 172.0/ 1.0 &  2.5 & 0.09 &  2.6 \nl

B & \ 0.83/0.06 & 0.04 & -0.24/0.04 & 0.05 & 0.87/0.05 & 0.03 & 171.9/ 1.5 &  1.7 & 0.05 &  1.8 \nl

V & \ 0.93/0.04 & 0.07 & -0.25/0.06 & 0.05 & 0.97/0.04 & 0.06 & 172.4/ 1.7 &  1.8 & 0.07 &  2.3 \nl

R & \ 0.91/0.05 & 0.04 & -0.24/0.04 & 0.05 & 0.95/0.04 & 0.04 & 172.6/ 1.5 &  1.8 & 0.07 &  1.6 \nl

I & \ 0.82/0.07 & 0.08 & -0.20/0.04 & 0.06 & 0.85/0.07 & 0.07 & 172.8/ 1.7 &  2.4 & 0.09 &  2.8 \nl

GAV & \ \ \ \ \ \ 0.06 \ & 0.06 \ & \ \ \ \ \ \ 0.04 \ & 0.06 \ &
\multicolumn{1}{r}{0.06} & 0.05 & \multicolumn{1}{r}{1.5} &  2.0 & 0.07 &  2.2 \nl

\sidehead{$\zeta$ Tau}

U & \ 0.51/0.06 & 0.05 & \ 1.06/0.05 & 0.06 & 1.19/0.06 & 0.05 & \ 32.2/ 1.4 &  1.4 &  0.07 &  2.0 \nl

B & \ 0.65/0.07 & 0.05 & \ 1.36/0.09 & 0.09 & 1.51/0.10 & 0.08 & \ 32.2/ 1.0 &  1.0 &  0.05 &  1.8 \nl

V & \ 0.63/0.06 & 0.05 & \ 1.29/0.06 & 0.07 & 1.44/0.08 & 0.06 & \ 32.0/ 0.7 &  1.1 &  0.06 &  1.6 \nl

R & \ 0.52/0.03 & 0.04 & \ 1.23/0.07 & 0.05 & 1.34/0.07 & 0.05 & \ 33.5/ 0.2 &  0.9 &  0.08 &  1.6 \nl

I & \ 0.46/0.11 & 0.04 & \ 1.06/0.05 & 0.11 & 1.16/0.07 & 0.11 & \ 33.3/ 2.4 &  1.0 &  0.08 &  2.0 \nl

GAV & \ \ \ \ \ \ 0.07 \ & 0.04 \ & \ \ \ \ \ \ 0.06 \ & 0.08 \ &
\multicolumn{1}{r}{0.08} & 0.07 & \multicolumn{1}{r}{1.1} &  1.1 & 0.07 &  1.8 \nl

\enddata
\end{deluxetable}

\begin{deluxetable}{lllllcccccc}
\scriptsize
\tt
\tablecolumns{11}
\tablewidth{0pt}
\tablecaption{Summer Be Star Summary}
\tablehead{
\colhead{Star} &
\colhead{$q\/$/$dq$} &
\colhead{<$dq$>} &
\colhead{$u\/$/$du$} &
\colhead{<$du$>} &
\colhead{$p\/$/$dp$} &
\colhead{<$dp$>} &
\colhead{$\theta\/$/$d\theta$} &
\colhead{<$d\theta$>} &
\colhead{<$dpi$>} &
\colhead{<$d\theta i$>} \nl
\colhead{Filter} &
\colhead{(\%)} &
\colhead{(\%)} &
\colhead{(\%)} &
\colhead{(\%)} &
\colhead{(\%)} &
\colhead{(\%)} &
\colhead{($\arcdeg$)} & 
\colhead{($\arcdeg$)} &
\colhead{(\%)} &
\colhead{($\arcdeg$)}
}

\startdata
\sidehead{48 Lib}

U & -0.38/0.04 & 0.08 & -0.56/0.12 & 0.08 & 0.69/0.11 & 0.08 & 117.6/ 1.9 &  3.6 & 0.12 &  5.1 \nl

B & -0.41/0.04 & 0.05 & -0.80/0.06 & 0.07 & 0.90/0.04 & 0.06 & 121.3/ 2.2 &  1.8 & 0.06 &  2.8 \nl

V & -0.52/0.06 & 0.04 & -0.70/0.03 & 0.06 & 0.88/0.02 & 0.07 & 116.7/ 2.1 &  1.2 & 0.07 &  2.2 \nl

R & -0.52/0.06 & 0.06 & -0.62/0.02 & 0.05 & 0.82/0.05 & 0.05 & 115.1/ 1.3 &  1.9 & 0.08 &  2.4 \nl

I & -0.53/0.10 & 0.05 & -0.56/0.04 & 0.10 & 0.79/0.08 & 0.09 & 113.3/ 2.6 &  3.1 & 0.12 &  5.6 \nl

GAV & \ \ \ \ \ \ 0.06 \ & 0.06 \ & \ \ \ \ \ \ 0.06 \ & 0.07 \ & 
\multicolumn{1}{r}{0.06} & 0.07 & \multicolumn{1}{r}{2.0} &  2.3 & 0.09 &  3.6 \nl

\sidehead{$\chi$ Oph}

U & \ 0.04/0.06 & 0.12 & -0.33/0.06 & 0.05 & 0.39/0.02 & 0.06 & 138.0/ 4.7 & 10.2 & 0.11 & 13.4 \nl

B & -0.02/0.03 & 0.08 & -0.46/0.04 & 0.06 & 0.48/0.04 & 0.06 & 134.6/ 1.7 &  5.8 & 0.09 &  5.9 \nl

V & -0.00/0.02 & 0.09 & -0.47/0.04 & 0.06 & 0.49/0.03 & 0.06 & 135.0/ 1.2 &  5.5 & 0.10 &  5.7 \nl

R & -0.01/0.05 & 0.08 & -0.47/0.01 & 0.06 & 0.48/0.01 & 0.05 & 134.7/ 3.1 &  4.8 & 0.08 &  6.4 \nl

I & -0.05/0.07 & 0.09 & -0.42/0.03 & 0.08 & 0.46/0.04 & 0.06 & 133.0/ 4.7 &  7.0 & 0.12 &  9.8 \nl

GAV & \ \ \ \ \ \ 0.05 \ & 0.09 \ & \ \ \ \ \ \ 0.04 \ & 0.06 \ &
\multicolumn{1}{r}{0.03} & 0.06 & \multicolumn{1}{r}{3.1} &  6.7 & 0.10 &  8.2 \nl

\sidehead{$\pi$ Aqr}

U & -0.25/0.16 & 0.06 & -0.33/0.08 & 0.09 & 0.46/0.04 & 0.09 & 116.1/10.6 &  4.2 & 0.11 &  7.9 \nl

B & -0.30/0.14 & 0.05 & -0.37/0.09 & 0.04 & 0.50/0.01 & 0.04 & 115.6/ 9.7 &  2.8 & 0.08 &  4.0 \nl

V & -0.32/0.17 & 0.06 & -0.35/0.09 & 0.04 & 0.51/0.01 & 0.03 & 114.3/11.1 &  3.7 & 0.07 &  3.9 \nl

R & -0.30/0.17 & 0.06 & -0.33/0.08 & 0.04 & 0.48/0.02 & 0.05 & 114.6/11.7 &  3.0 & 0.08 &  4.1 \nl

I & -0.22/0.18 & 0.09 & -0.31/0.07 & 0.05 & 0.44/0.02 & 0.07 & 118.3/13.1 &  6.1 & 0.09 &  9.7 \nl

GAV & \ \ \ \ \ \ 0.16 \ & 0.06 \ & \ \ \ \ \ \ 0.08 \ & 0.05 \ &
\multicolumn{1}{r}{0.02} & 0.05 & \multicolumn{1}{r}{11.2} &  4.0 & 0.09 &  5.9 \nl

\sidehead{o And}

U & -0.39/0.09 & 0.05 & -0.11/0.11 & 0.05 & 0.42/0.11 & 0.04 & \ 96.3/ 6.6 &  3.9 & 0.08 &  5.7 \nl

B & -0.46/0.12 & 0.04 & -0.13/0.11 & 0.06 & 0.49/0.14 & 0.05 & \ 97.1/ 4.3 &  3.6 & 0.06 &  3.9 \nl

V & -0.45/0.13 & 0.05 & -0.08/0.07 & 0.04 & 0.47/0.14 & 0.05 & \ 94.8/ 3.6 &  2.6 & 0.06 &  5.4 \nl

R & -0.43/0.13 & 0.05 & -0.08/0.07 & 0.04 & 0.45/0.14 & 0.05 & \ 94.8/ 3.1 &  2.4 & 0.05 &  5.3 \nl

I & -0.40/0.13 & 0.08 & -0.06/0.09 & 0.07 & 0.43/0.14 & 0.09 & \ 93.8/ 4.6 &  5.1 & 0.08 &  7.7 \nl

GAV & \ \ \ \ \ \ 0.12 \ & 0.05 \ & \ \ \ \ \ \ 0.09 \ & 0.05 \ &
\multicolumn{1}{r}{0.13} & 0.05 & \multicolumn{1}{r}{4.5} &  3.5 & 0.07 &  5.6 \nl

\enddata
\end{deluxetable}

\begin{deluxetable}{lllllcccccc}
\scriptsize
\tt
\tablecolumns{11}
\tablewidth{0pt}
\tablecaption{Polarized Standard Star Summary (Limber)}
\tablehead{
\colhead{Star} &
\colhead{$q\/$/$dq$} &
\colhead{<$dq$>} &
\colhead{$u\/$/$du$} &
\colhead{<$du$>} &
\colhead{$p\/$/$dp$} &
\colhead{<$dp$>} &
\colhead{$\theta\/$/$d\theta$} &
\colhead{<$d\theta$>} &
\colhead{<$dpi$>} &
\colhead{<$d\theta i$>} \nl
\colhead{Filter} &
\colhead{(\%)} &
\colhead{(\%)} &
\colhead{(\%)} &
\colhead{(\%)} &
\colhead{(\%)} &
\colhead{(\%)} &
\colhead{($\arcdeg$)} & 
\colhead{($\arcdeg$)} &
\colhead{(\%)} &
\colhead{($\arcdeg$)}
}

\startdata
\sidehead{2H Cam}

U & -1.83/0.09 & 0.04 & -2.41/0.06 & 0.04 & 3.03/0.07 & 0.03 & 116.5/ 0.8 & 0.5& 0.09 & 1.1 \nl

B & -1.95/0.08 & 0.07 & -2.65/0.06 & 0.05 & 3.29/0.02 & 0.06 & 116.8/ 0.9 & 0.6& 0.06 & 0.6 \nl

V & -2.05/0.02 & 0.06 & -2.75/0.04 & 0.06 & 3.43/0.02 & 0.06 & 116.7/ 0.4 & 0.6& 0.06 & 0.6 \nl

R & -1.98/0.02 & 0.05 & -2.66/0.04 & 0.06 & 3.32/0.02 & 0.05 & 116.7/ 0.3 & 0.6& 0.05 & 0.5 \nl

I & -1.79/0.03 & 0.07 & -2.37/0.09 & 0.06 & 2.96/0.06 & 0.08 & 116.4/ 0.6 & 0.7& 0.09 & 0.8 \nl

GAV & \ \ \ \ \ \ 0.05 \ & 0.06 \ & \ \ \ \ \ \ 0.06 \ & 0.06 \ &
\multicolumn{1}{r}{0.04} & 0.06 & \multicolumn{1}{r}{0.6} & 0.6 & 0.07 & 0.7 \nl

\sidehead{o Sco}

U & \ 1.06/0.14 & 0.18 & \ 2.35/0.05 & 0.23 & 2.59/0.10 & 0.25 & \ 33.0/ 1.3 &  1.7 & 0.19 & 2.7 \nl

B & \ 1.41/0.11 & 0.07 & \ 3.09/0.06 & 0.06 & 3.40/0.04 & 0.08 & \ 32.8/ 1.0 &  0.5 & 0.08 & 0.8 \nl

V & \ 1.73/0.16 & 0.08 & \ 3.76/0.05 & 0.08 & 4.15/0.05 & 0.08 & \ 32.7/ 1.1 &  0.5 & 0.10 & 0.8 \nl

R & \ 1.83/0.16 & 0.07 & \ 4.00/0.06 & 0.07 & 4.40/0.04 & 0.08 & \ 32.8/ 1.1 &  0.4 & 0.09 & 0.6 \nl

I & \ 1.74/0.15 & 0.07 & \ 3.93/0.10 & 0.05 & 4.30/0.05 & 0.05 & \ 33.1/ 1.2 &  0.5 & 0.14 & 1.0 \nl

GAV & \ \ \ \ \ \ 0.14 \ & 0.10 \ & \ \ \ \ \ \ 0.06 \ & 0.10 \ &
\multicolumn{1}{r}{0.06} & 0.11 & \multicolumn{1}{r}{1.1} & 0.7 & 0.12 & 1.2 \nl

\enddata
\end{deluxetable}

\begin{deluxetable}{lllllcccccc}
\scriptsize
\tt
\tablecolumns{11}
\tablewidth{0pt}
\tablecaption{Polarized Standard Star Summary (McDonald)}
\tablehead{
\colhead{Star} &
\colhead{$q\/$/$dq$} &
\colhead{<$dq$>} &
\colhead{$u\/$/$du$} &
\colhead{<$du$>} &
\colhead{$p\/$/$dp$} &
\colhead{<$dp$>} &
\colhead{$\theta\/$/$d\theta$} &
\colhead{<$d\theta$>} &
\colhead{<$dpi$>} &
\colhead{<$d\theta i$>} \nl
\colhead{Filter} &
\colhead{(\%)} &
\colhead{(\%)} &
\colhead{(\%)} &
\colhead{(\%)} &
\colhead{(\%)} &
\colhead{(\%)} &
\colhead{($\arcdeg$)} & 
\colhead{($\arcdeg$)} &
\colhead{(\%)} &
\colhead{($\arcdeg$)}
}

\startdata
\sidehead{2H Cam}

U & -1.83/0.05 \ & 0.11 \ & -2.41/0.03 \ & 0.07 \ & 3.03/0.01 & 0.08 & 116.5/ 0.6 &  1.1 &  0.09 &  0.8 \nl

B & -1.97/0.05 \ & 0.07 \ & -2.60/0.09 \ & 0.07 \ & 3.26/0.07 & 0.05 & 116.3/ 0.6 &  0.7 &  0.04 &  0.4 \nl

V & -2.07/0.06 \ & 0.07 \ & -2.73/0.01 \ & 0.07 \ & 3.43/0.03 & 0.07 & 116.4/ 0.5 &  0.5 &  0.04 &  0.4 \nl

R & -1.97/0.07 \ & 0.05 \ & -2.66/0.05 \ & 0.06 \ & 3.31/0.04 & 0.05 & 116.8/ 0.7 &  0.4 &  0.03 &  0.3 \nl

I & -1.75/0.06 \ & 0.09 \ & -2.36/0.05 \ & 0.06 \ & 2.94/0.05 & 0.08 & 116.8/ 0.7 &  0.8 &  0.05 &  0.6 \nl

GAV & \ \ \ \ \ \ 0.06 \ & 0.08 \ & \ \ \ \ \ \ 0.05 \ & 0.07 \ & 
\multicolumn{1}{r}{0.04} & 0.07 &  \multicolumn{1}{r}{0.6} &  0.7 & 0.05 &  0.5 
\nl
\sidehead{o Sco}

U & \ 1.35/0.21 \ & 0.20 \ & \ 2.51/0.12 \ & 0.25 \ & 2.85/0.12 & 0.28 & \ 30.9/ 2.0 &  1.6 &  0.20 &  2.1 \nl

B & \ 1.40/0.12 \ & 0.09 \ & \ 3.07/0.14 \ & 0.08 \ & 3.39/0.10 & 0.07 & \ 32.7/ 1.3 &  0.9 &  0.06 &  0.5 \nl

V & \ 1.81/0.25 \ & 0.12 \ & \ 3.74/0.12 \ & 0.08 \ & 4.16/0.01 & 0.10 & \ 32.1/ 1.9 &  0.7 &  0.05 &  0.3 \nl

R & \ 1.91/0.23 \ & 0.07 \ & \ 3.96/0.10 \ & 0.08 \ & 4.41/0.03 & 0.05 & \ 32.1/ 1.6 &  0.6 &  0.04 &  0.2 \nl

I & \ 1.76/0.19 \ & 0.07 \ & \ 3.88/0.06 \ & 0.06 \ & 4.26/0.06 & 0.07 & \ 32.8/ 1.3 &  0.4 &  0.05 &  0.3 \nl

GAV & \ \ \ \ \ \ 0.20 \ & 0.11 \ & \ \ \ \ \ \ 0.11 \ & 0.11 \ & \multicolumn{1
}{r}{0.06} & 0.11 &  \multicolumn{1}{r}{1.6} &  0.9 & 0.08 &  0.7 \nl

\enddata
\end{deluxetable}

\clearpage

\begin{deluxetable}{lcccl}
\tt
\tablecolumns{5}
\tablewidth{0pt}
\tablecaption{Interstellar Polarization Parameters}
\tablehead{
\colhead{Star} &
\colhead{\ $p_{max}$} &
\colhead{\ $\lambda_{max}$} &
\colhead{\ $\theta_{IS}$} &
\colhead{Reference} \nl
\ & (\%) & (\AA) & ($\arcdeg$) & \
}

\startdata

$\gamma$ Cas & 0.26 & 5900 & \ 95 & MB \nl

$\phi$ Per & 1.06 & 4460 & 106 & PBL \nl

48 Per & 0.85 & 5990 & \ \ 0 & PBL \nl

$\zeta$ Tau & 0.00 & \ldots & \ldots & PBL \nl

48 Lib & 0.64 & 5670 & \ 85 & PBL \nl

$\chi$ Oph & 0.37 & 5740 & 158 & PBL \nl

$\pi$ Aqr & 0.46 & 4980 & 116 & PBL \nl

o And & 0.26 & 5970 & \ 75 & PBL \nl

\enddata
\end{deluxetable}

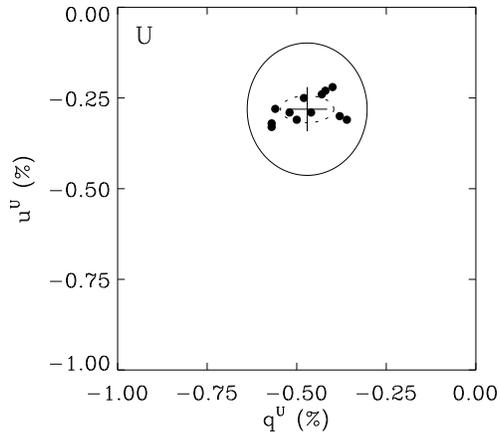
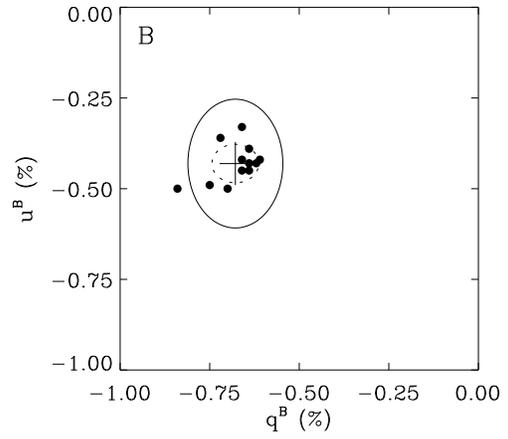
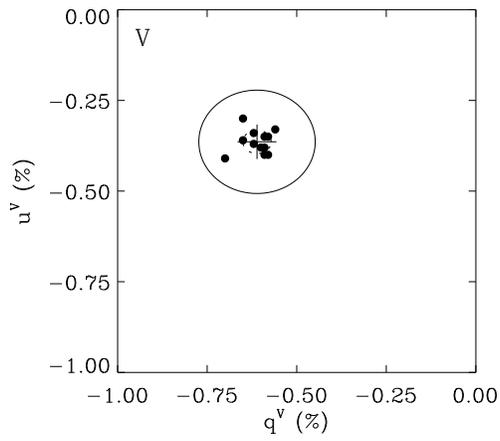
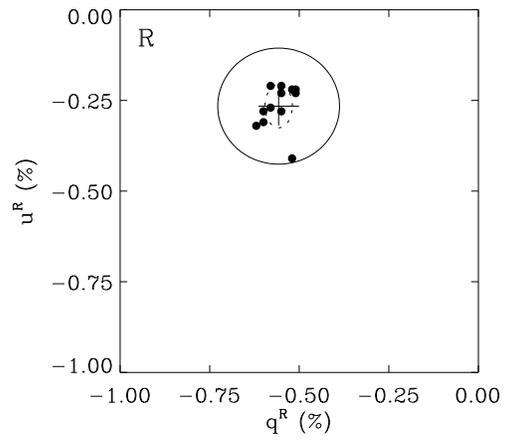
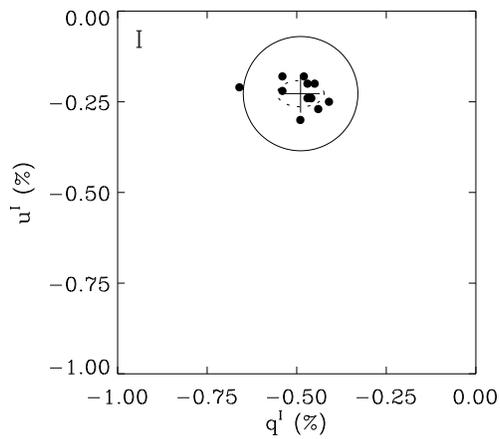
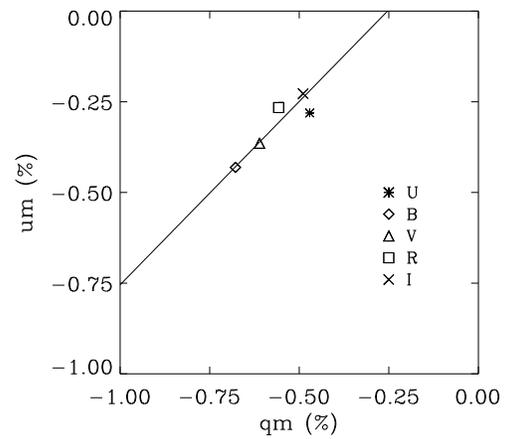
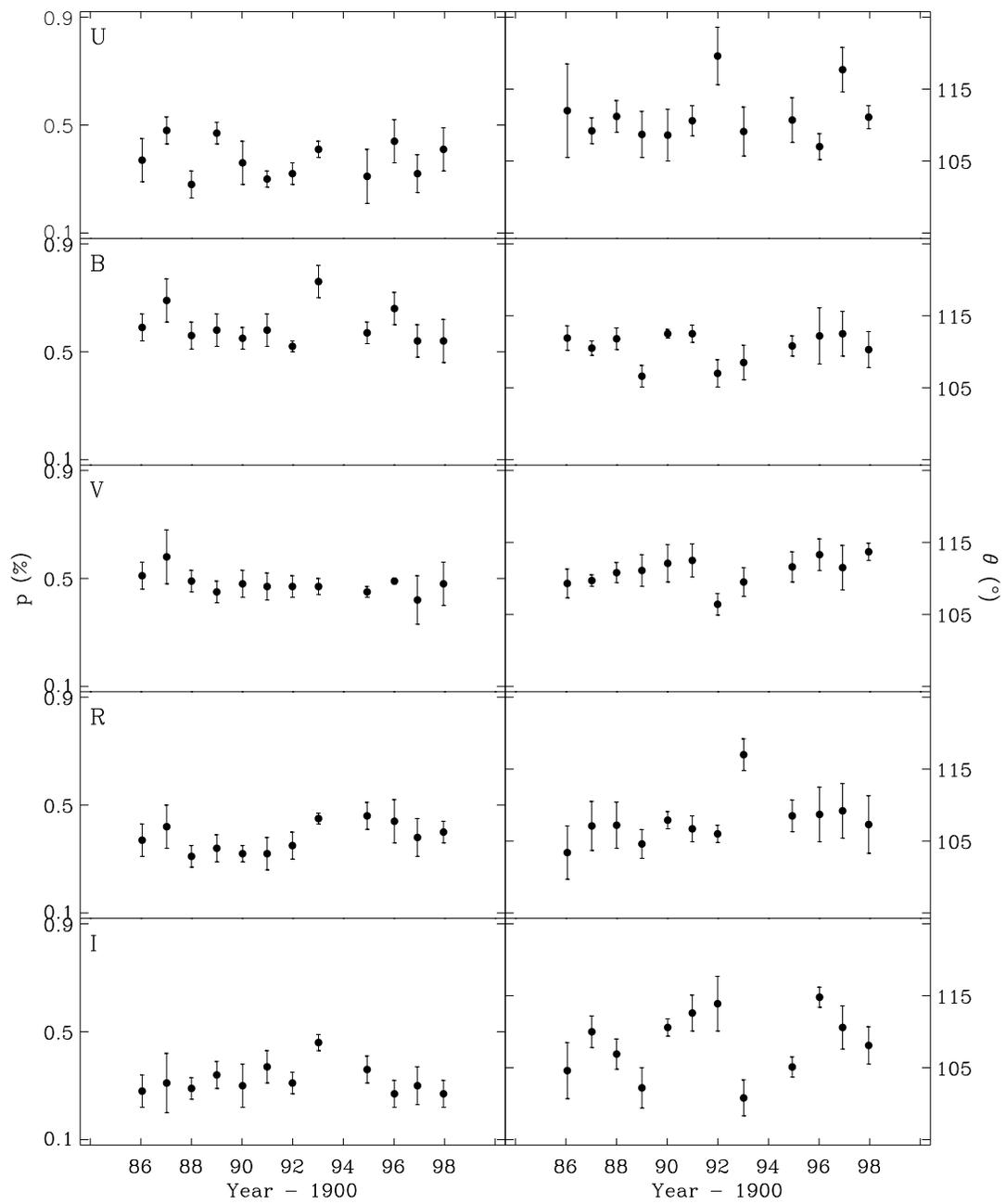
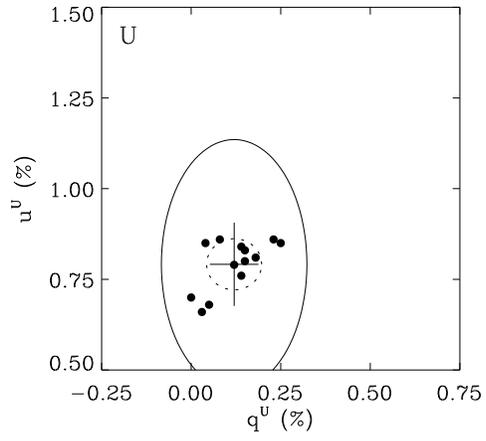
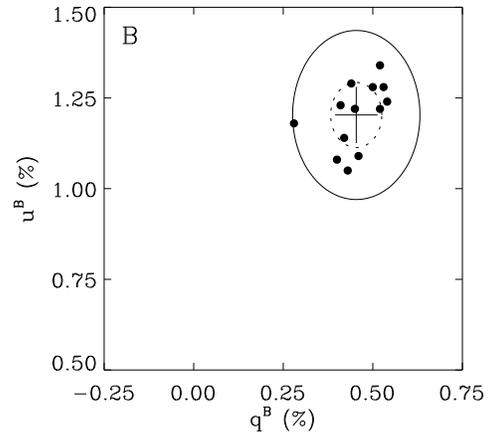
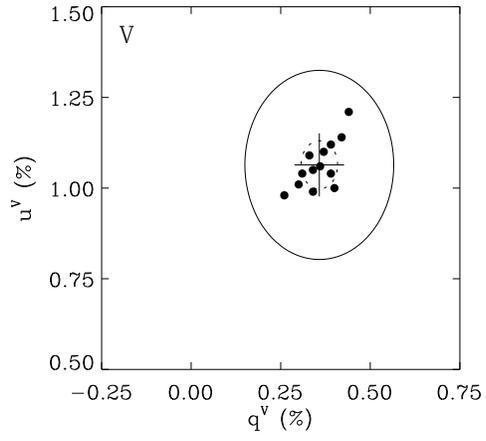
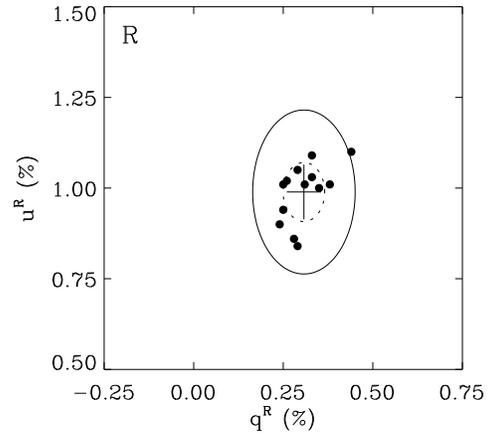
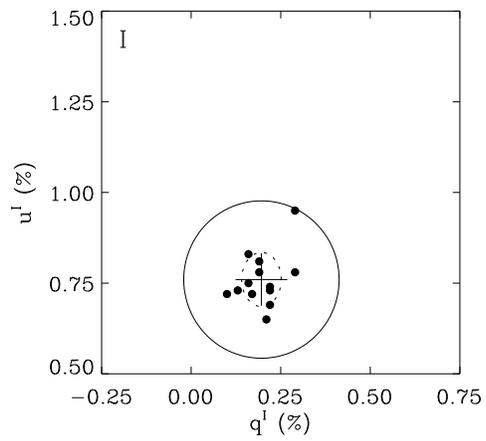
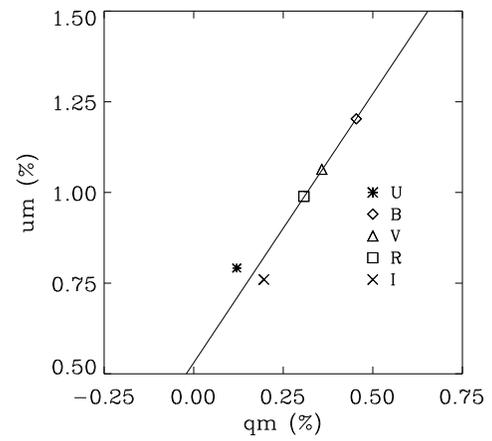
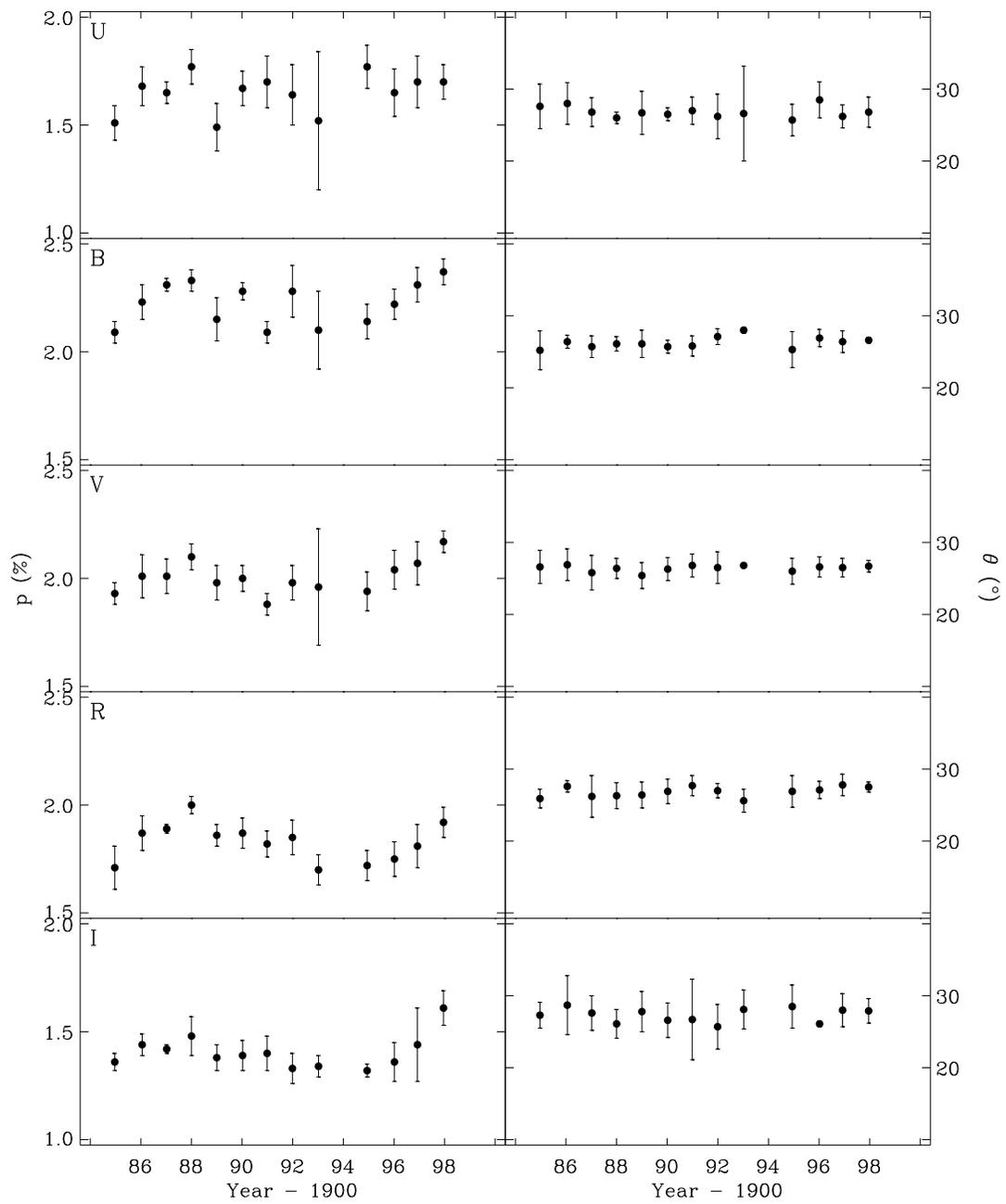
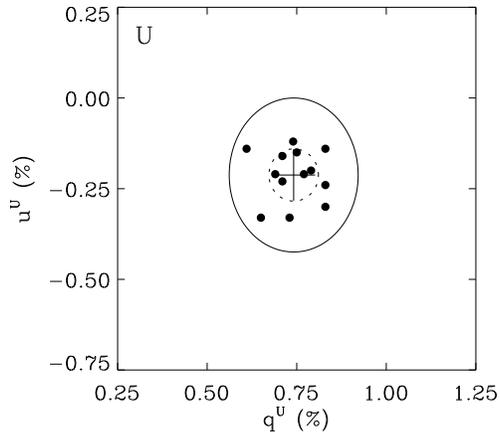
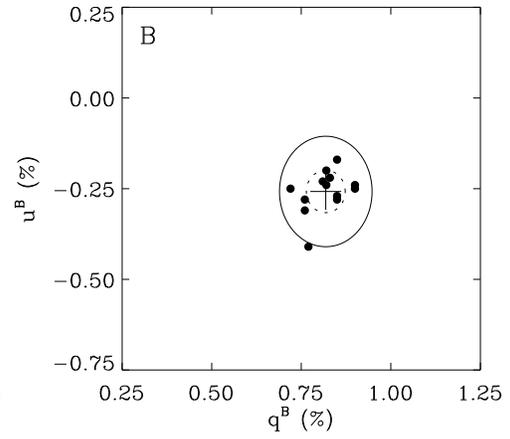
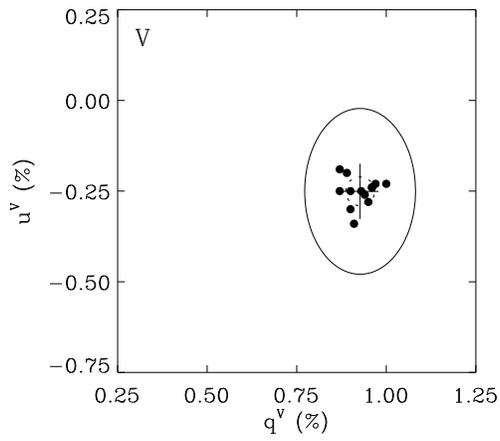
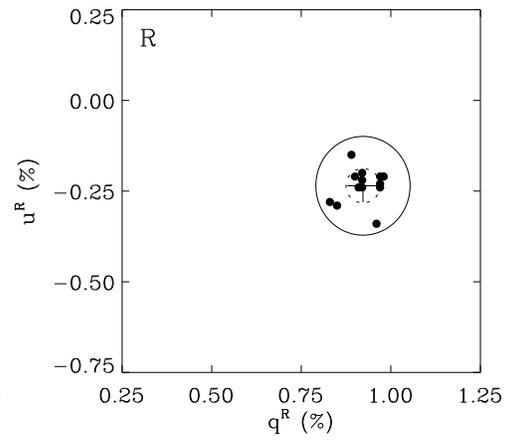
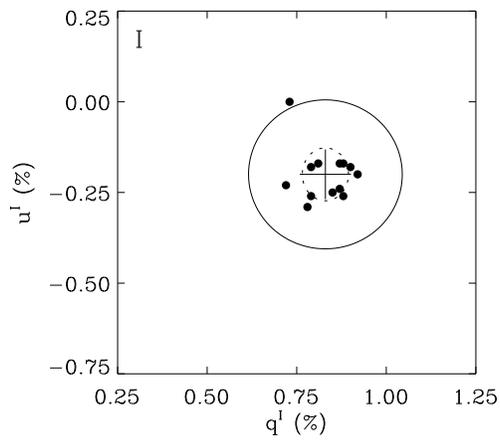
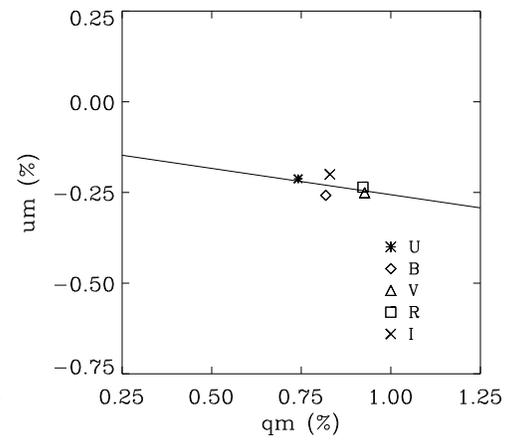
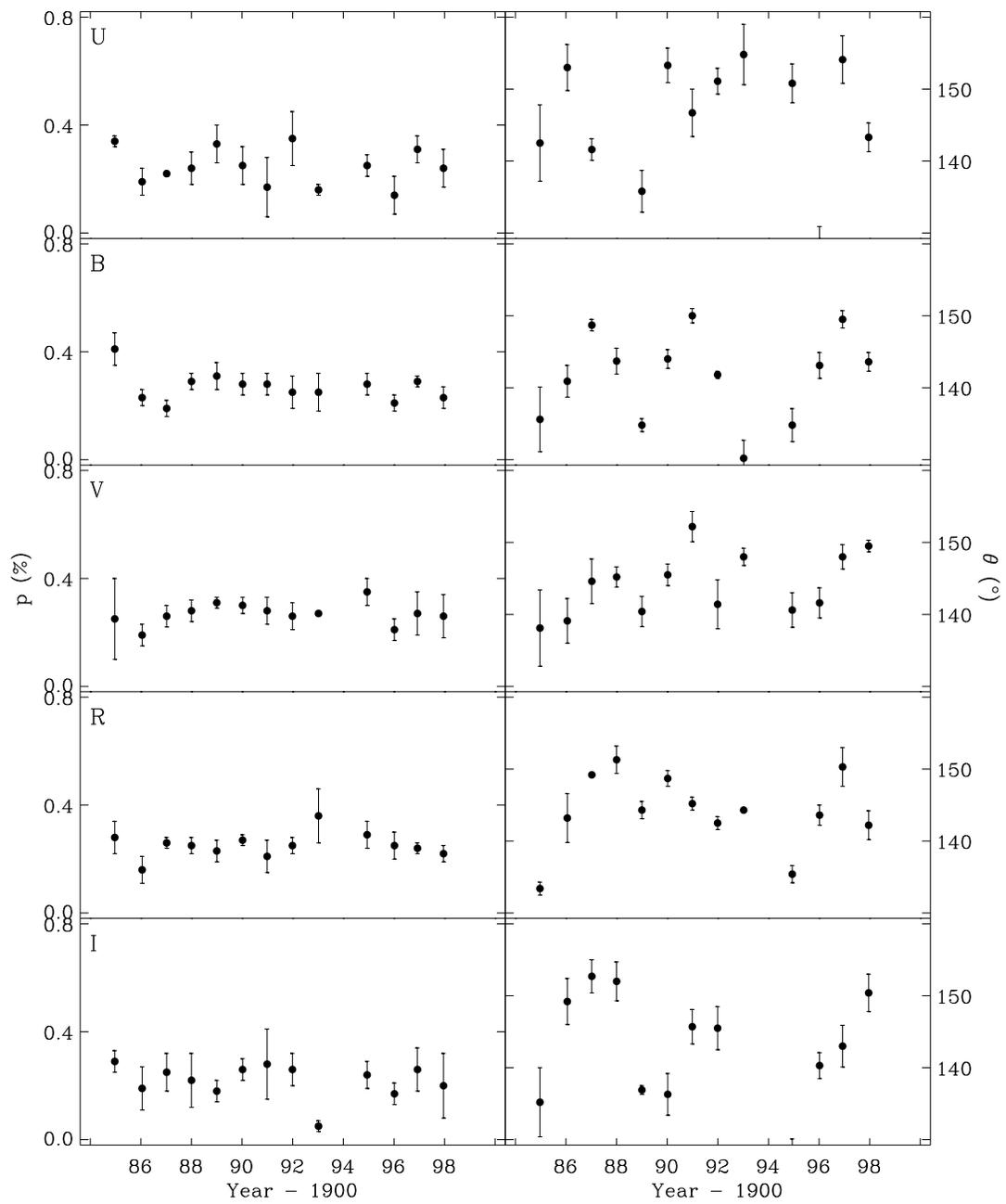
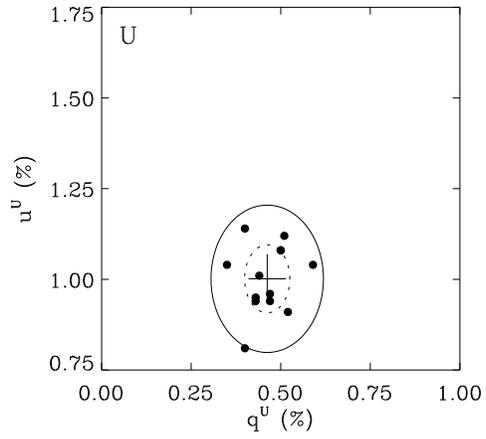
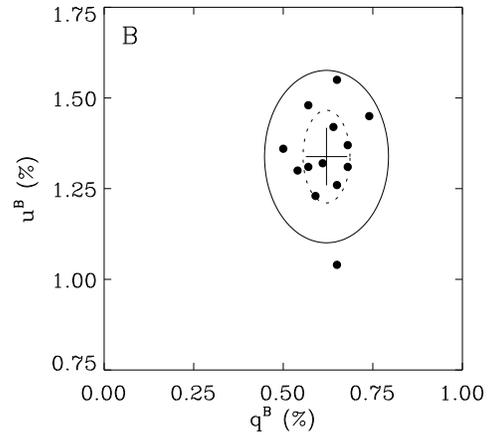
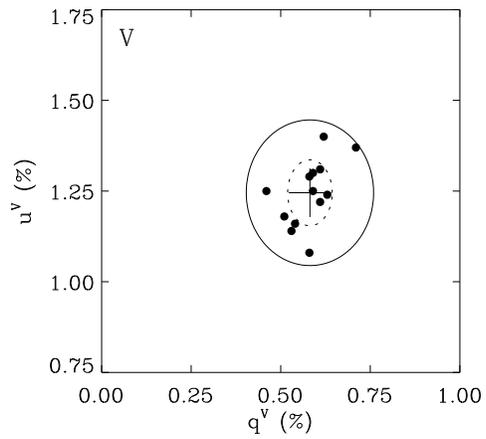
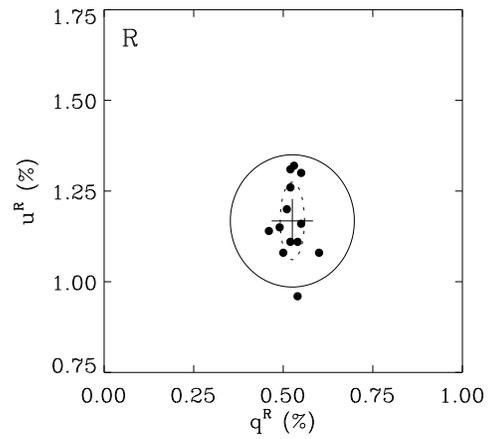
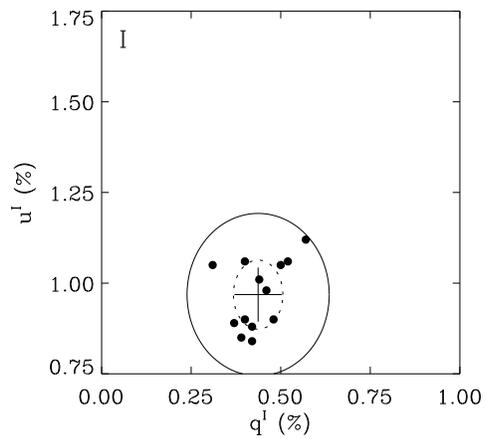
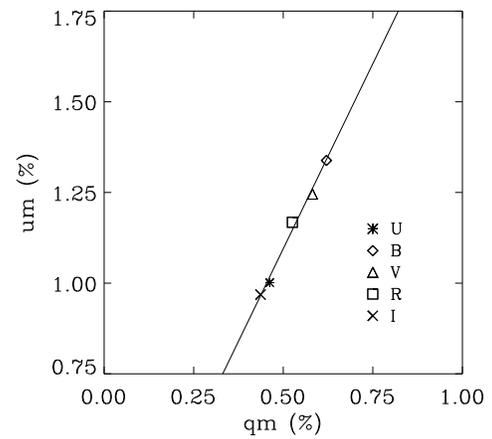
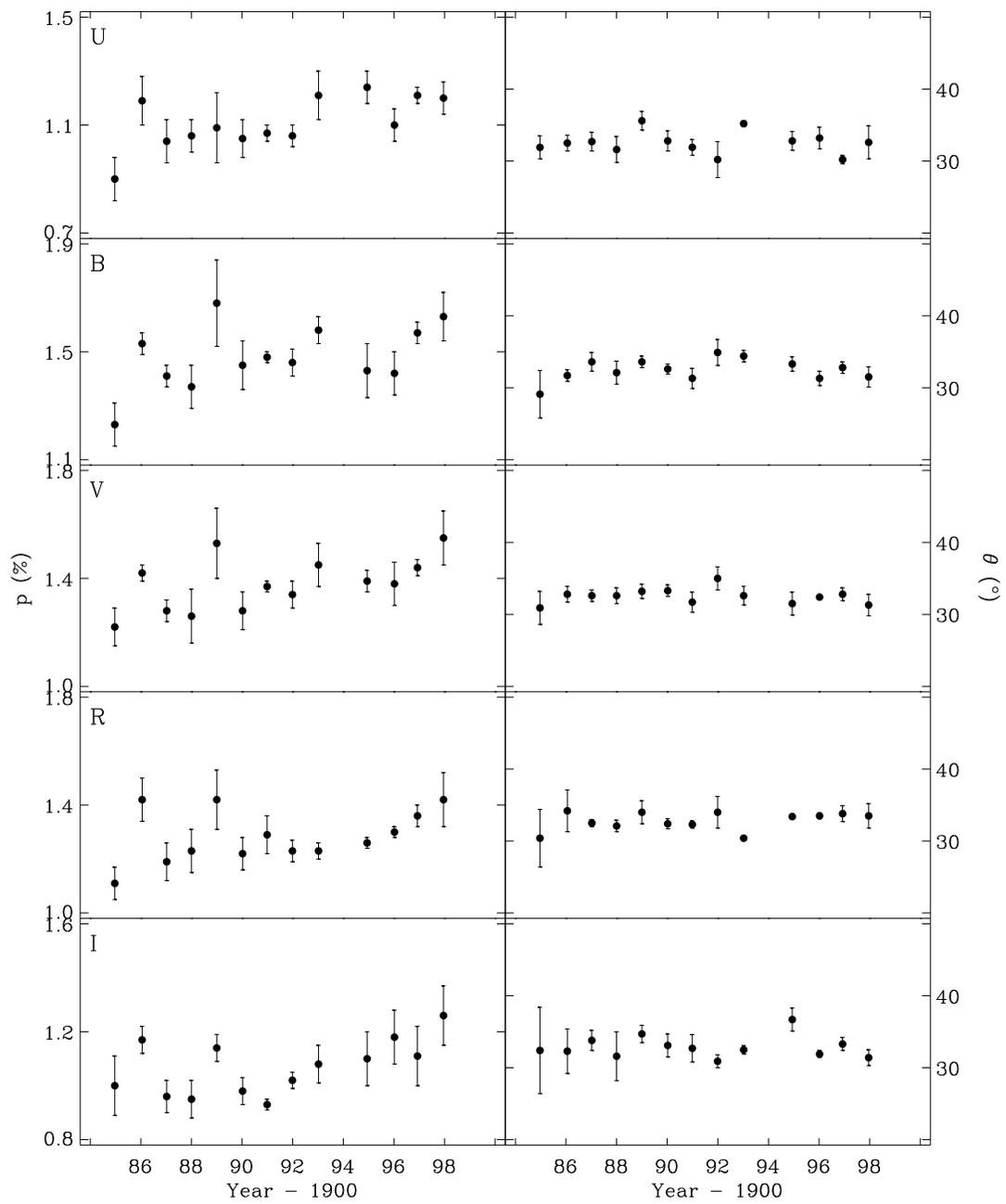
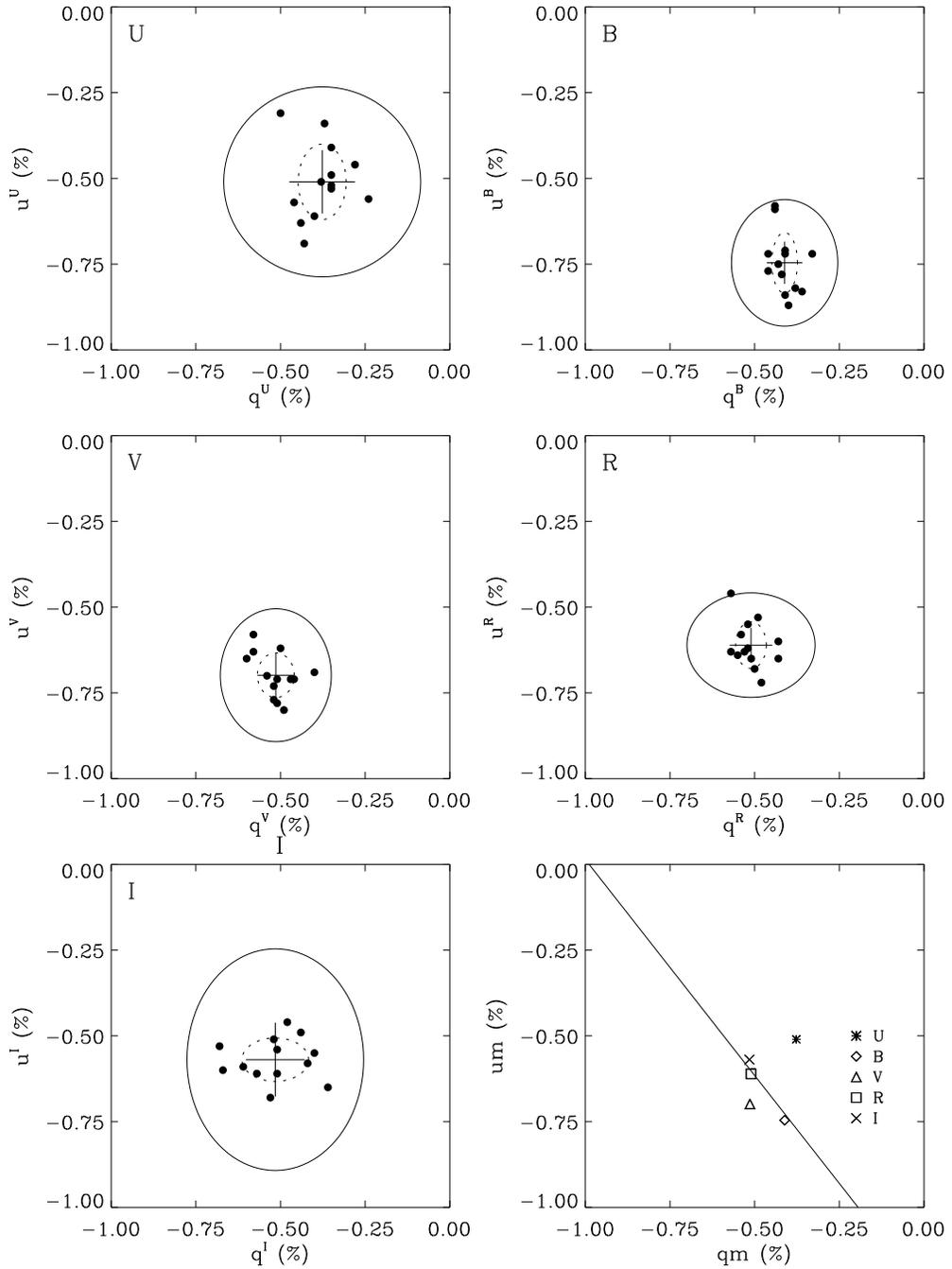
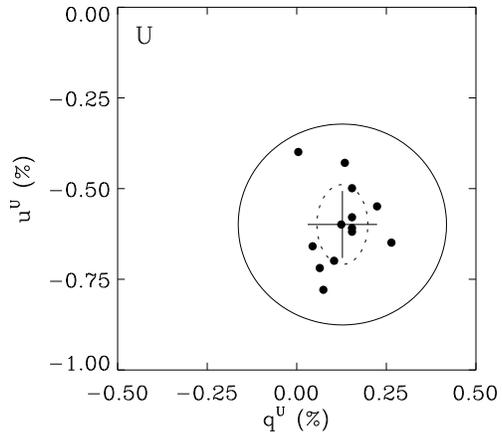
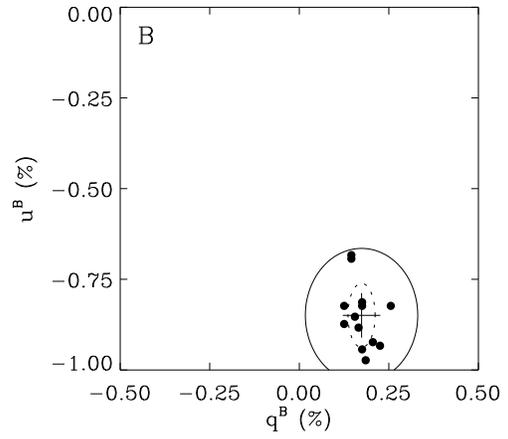
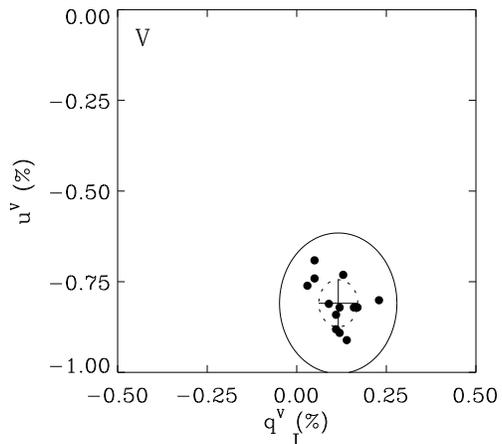
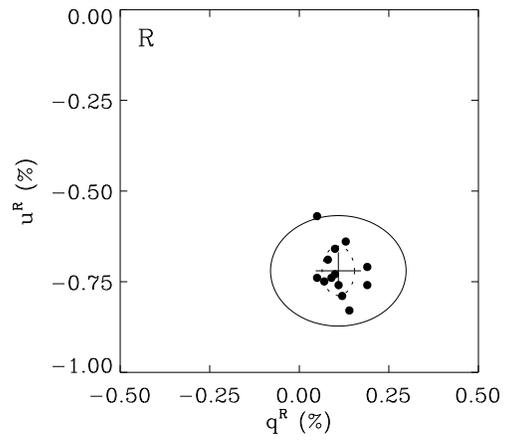
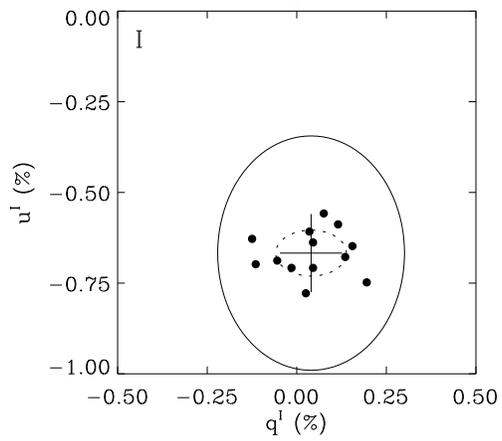
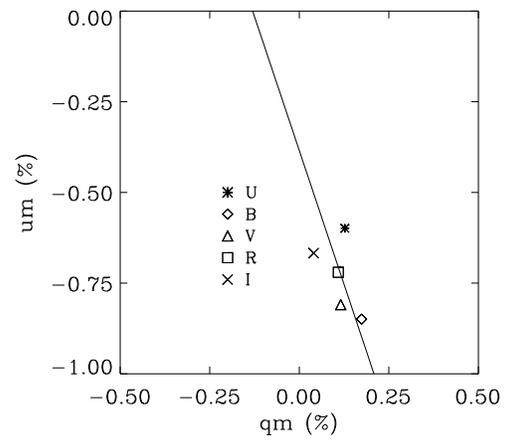
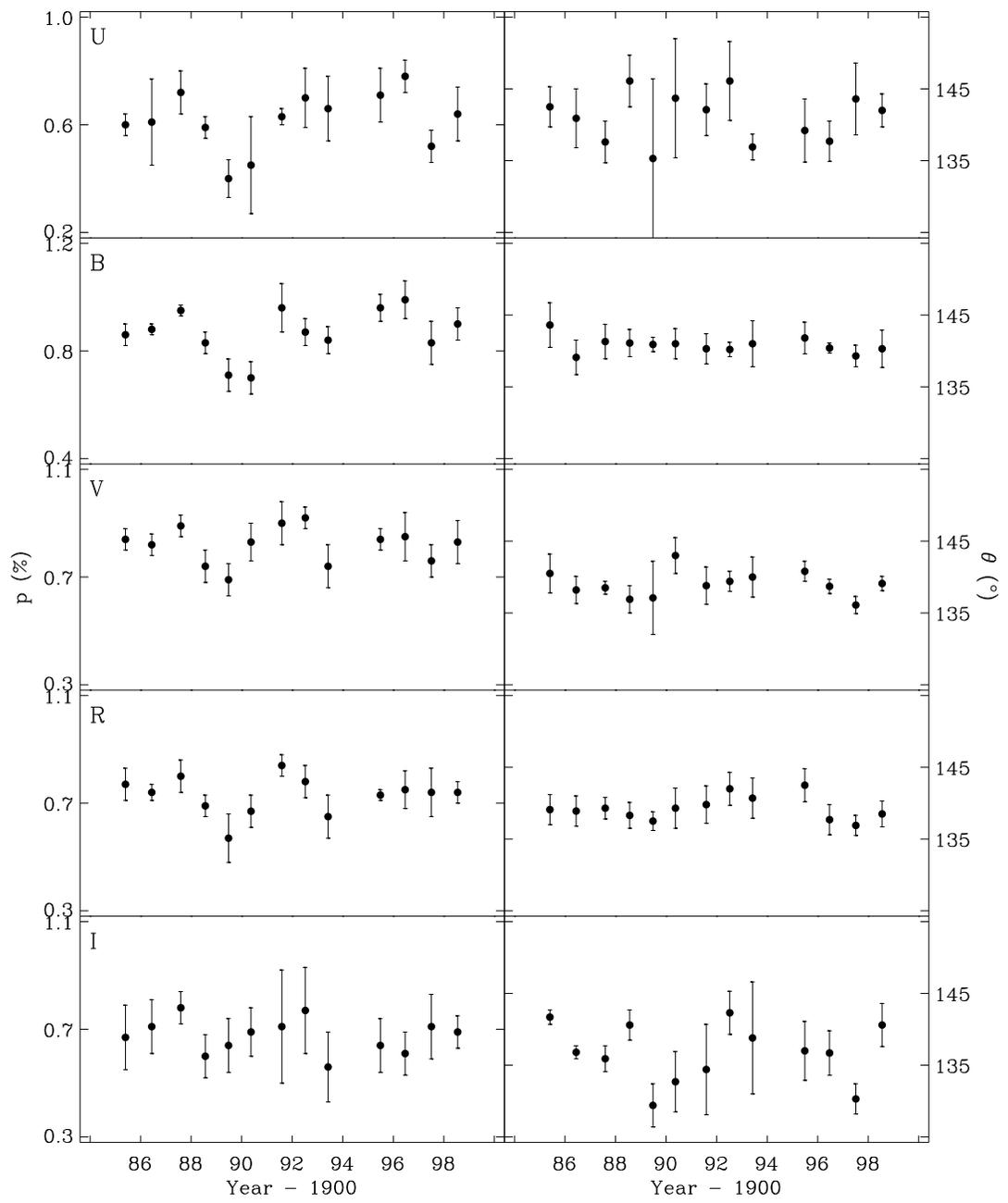
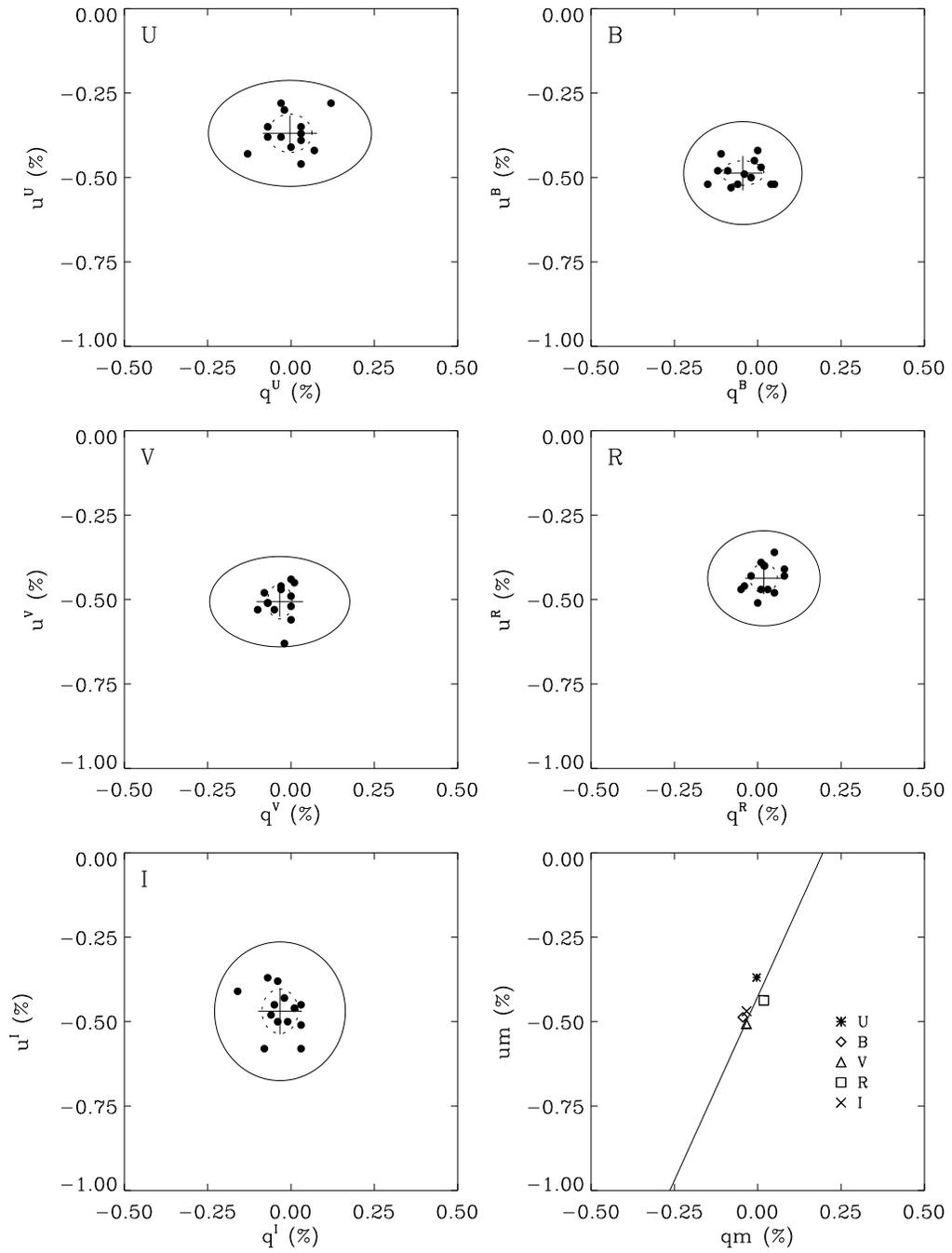
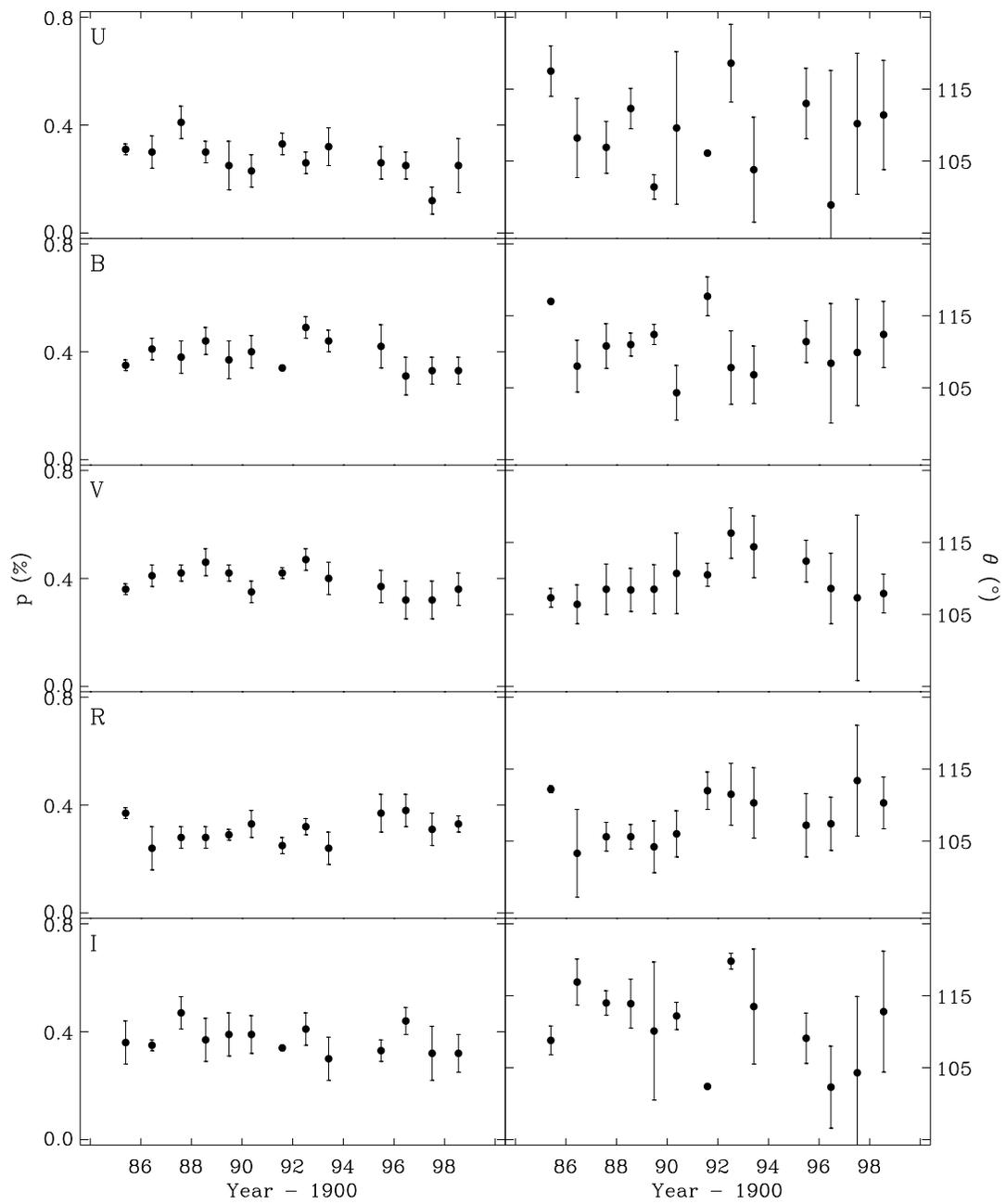
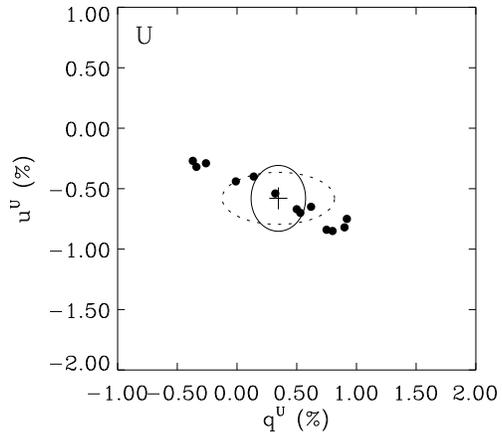
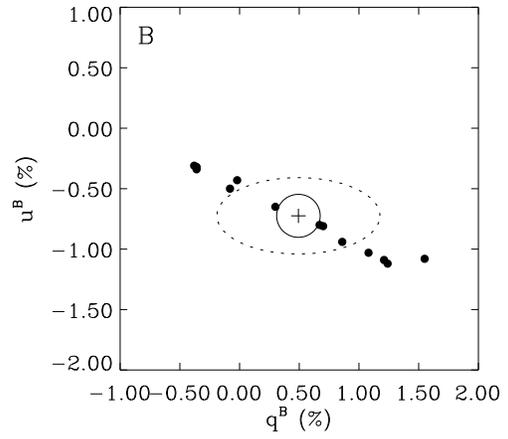
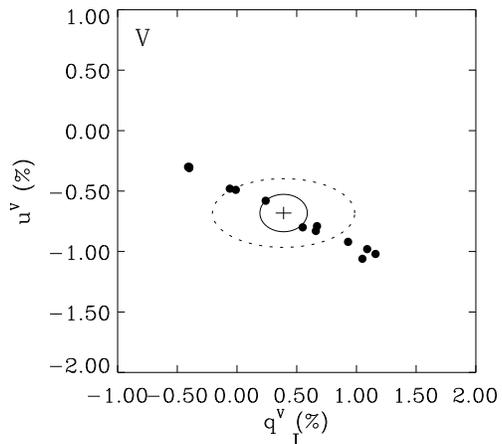
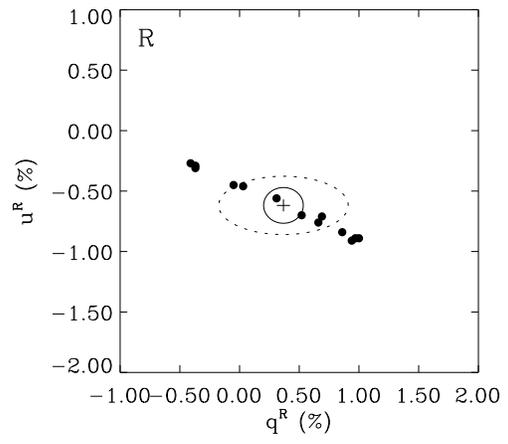
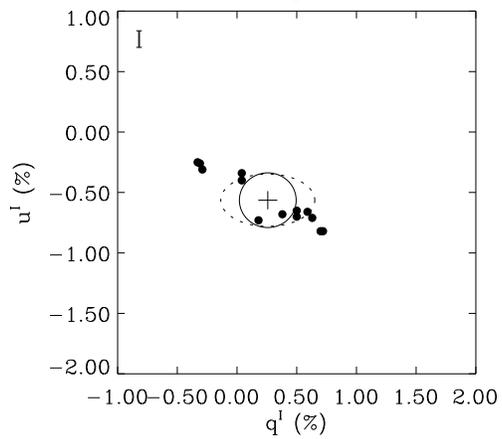
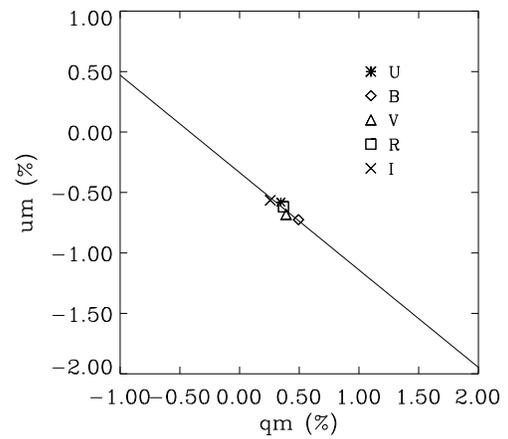
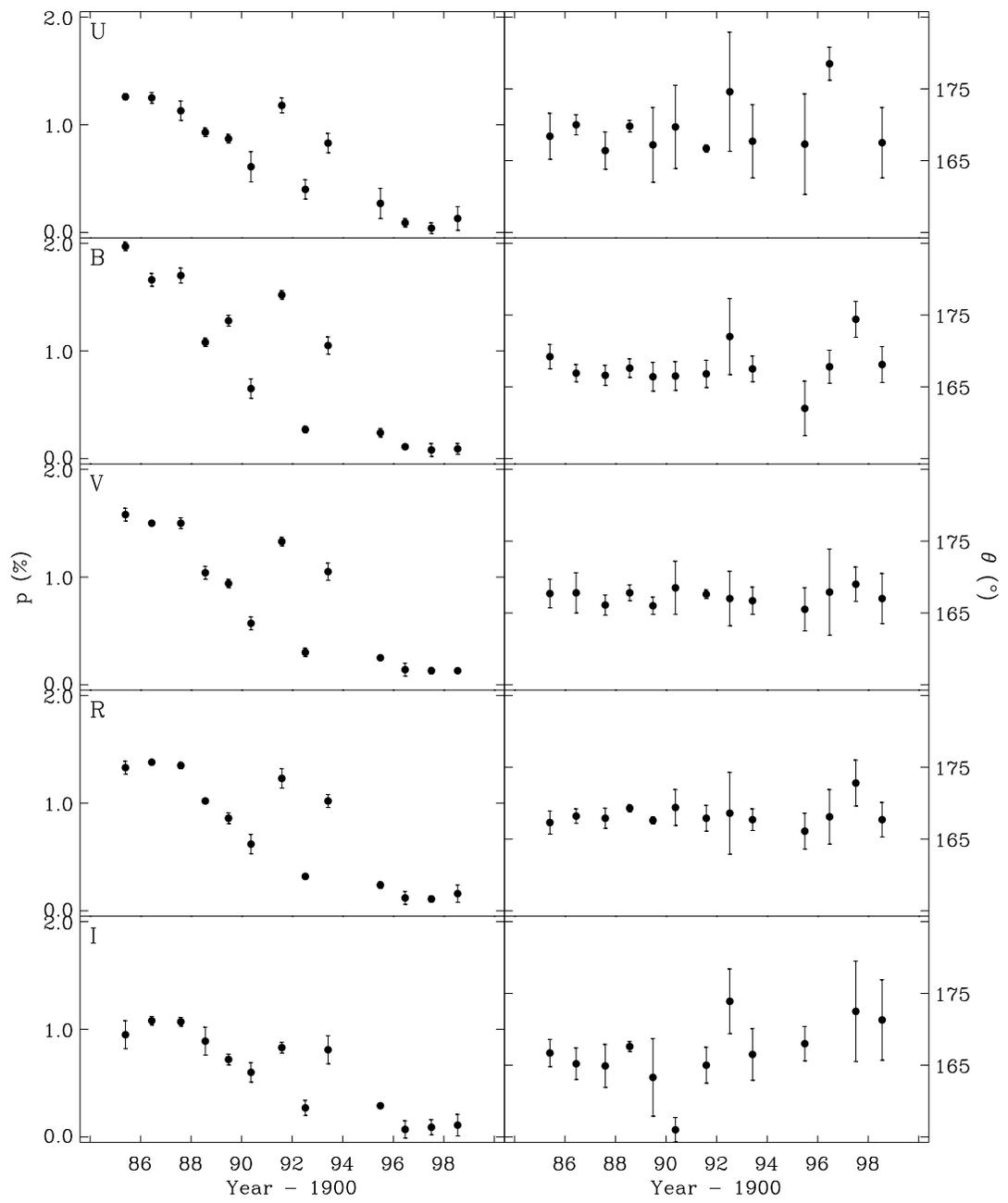
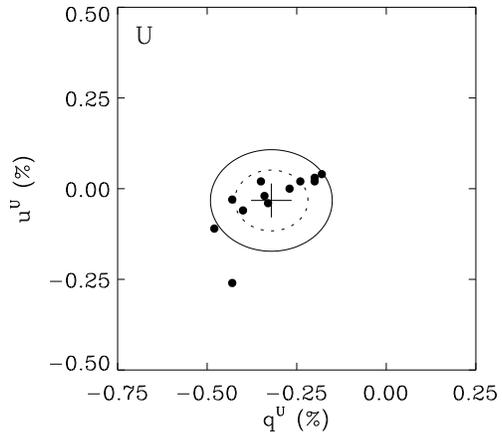
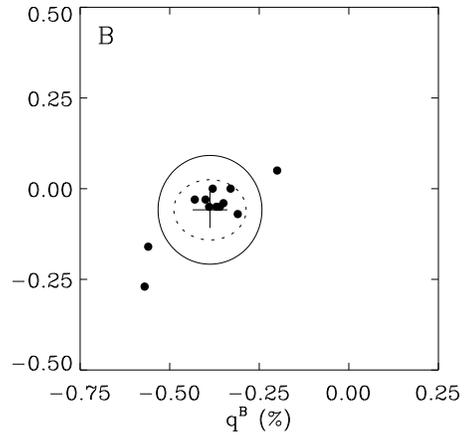
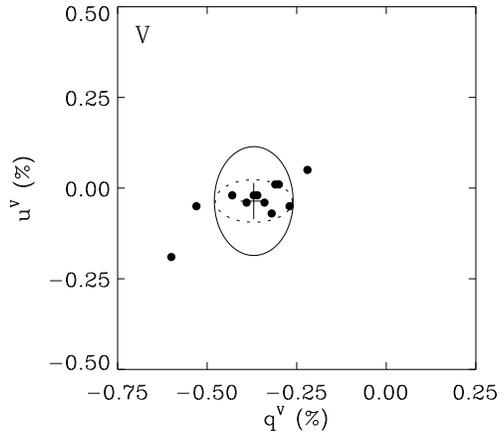
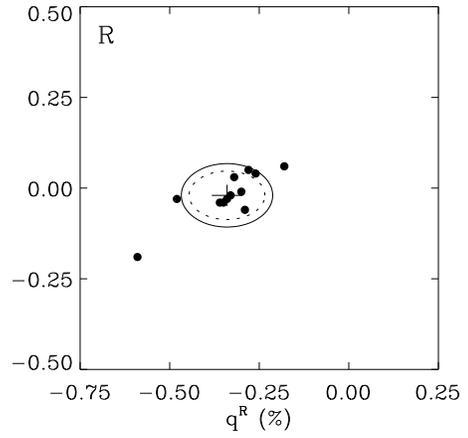
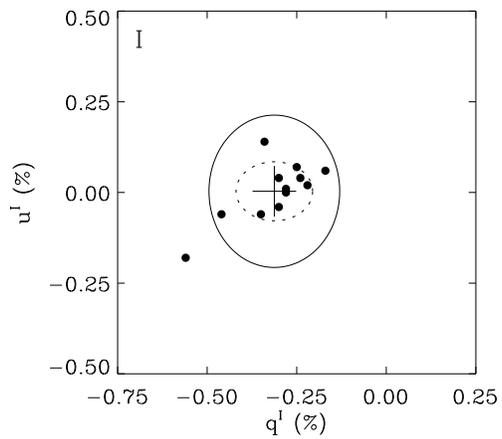
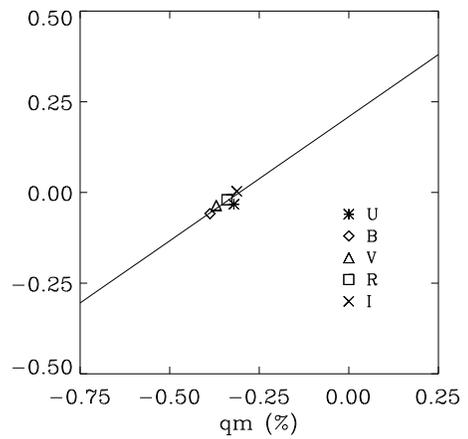
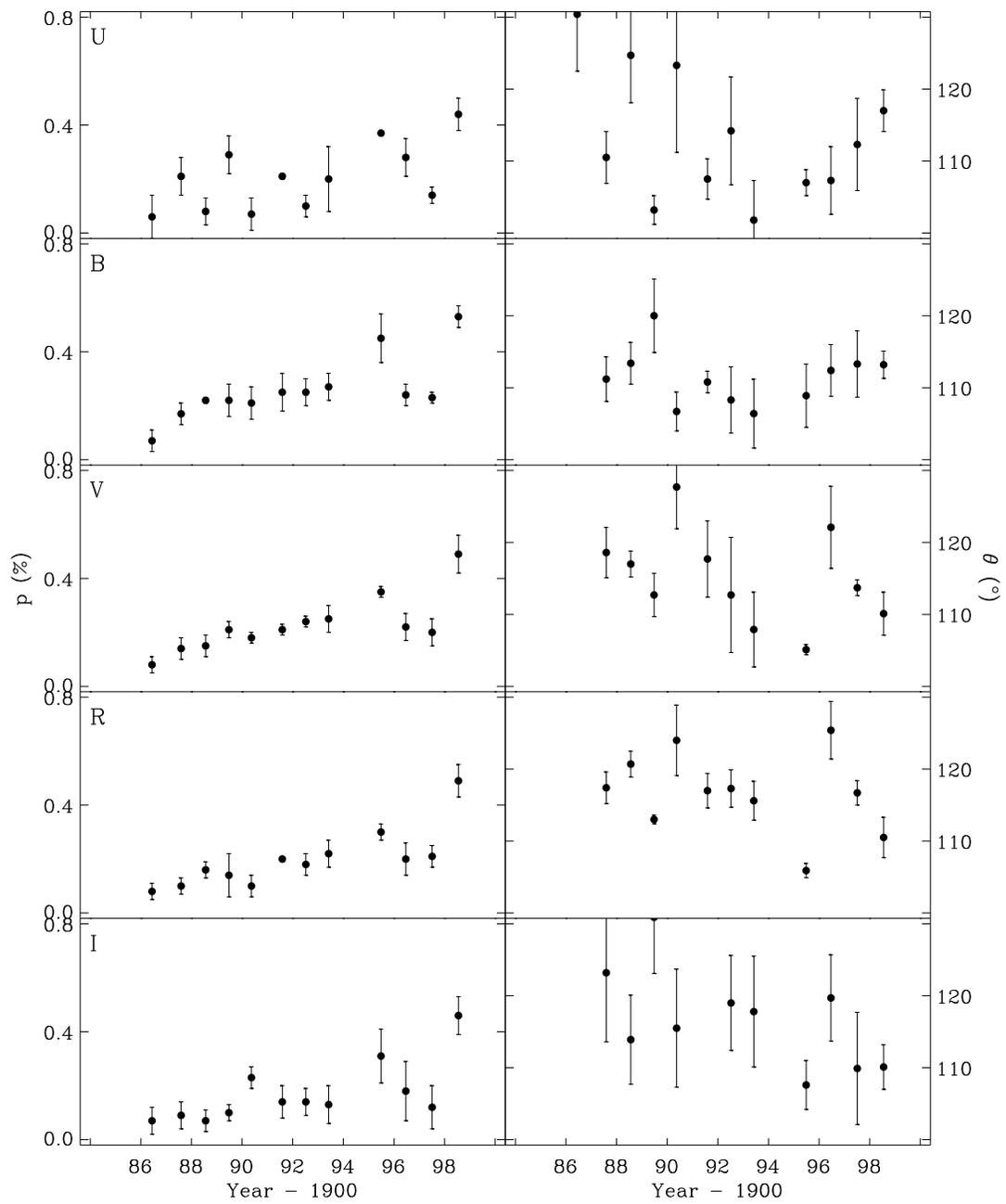
\end{document}